\documentclass[12pt]{article}
\usepackage[normalem]{ulem}
\usepackage{amsfonts,slashed}
\usepackage{arydshln}
\usepackage{fullpage}
\usepackage[greek,english]{babel}
\usepackage{cancel}
\usepackage{upgreek}
\usepackage{float}
\usepackage{url}
\usepackage{latexsym}
\usepackage{mathtools}
\usepackage{amsfonts}
\usepackage{amsmath}
\usepackage{epsfig}
\usepackage{latexsym,amssymb}
\usepackage{amsmath,amssymb,amsthm}
\usepackage{hyperref}
\usepackage{fontenc}
\usepackage{graphicx,tikz}
\usepackage{multirow}
\usepackage{xfrac}
\usepackage[T1]{fontenc}
\usepackage{textcomp}
\usepackage{lmodern}
\usepackage{stmaryrd}
\usepackage{cite}

\usepackage[margin=20pt,small]{caption}
\usepackage{subcaption}

\usepackage{ifpdf}
\ifx\pdfoutput\undefined
   \pdffalse
   \usepackage{cite}
 \else
   \pdfoutput=1
   \pdftrue
\fi
\DeclareFontFamily{OMS}{rsfs}{\skewchar\font'60}
\DeclareFontShape{OMS}{rsfs}{m}{n}{<-5>rsfs5 <5-7>rsfs7 <7->rsfs10 }{}
\DeclareSymbolFont{rsfs}{OMS}{rsfs}{m}{n}
\DeclareSymbolFontAlphabet{\Scr}{rsfs}

\numberwithin{equation}{section}
\def\be{\begin{equation}}
\def\ee{\end{equation}}
\def\ba{\begin{array}}
\def\ea{\end{array}}

\newcommand{\bea}{\begin{eqnarray}}
\newcommand{\eea}{\end{eqnarray}}

\newcommand{\mm}{m}

\begin{document}
\begin{titlepage}
%\vfill
%\begin{flushright}
%...
%\end{flushright}

%\vfill
\begin{center}
	{\LARGE \bf% Minimal massive supergravity and beyond -- \\[1ex]
%	New 3d (super-)gravities
	Minimal massive supergravity and\\ new theories of massive gravity\\[1cm]}
%\\~\\	Minimal massive supergravity\\ in a new family of massive gravities	}\\[1cm]
\hbox{{ \bf 
	\!\!\!\!\!Nihat Sadik Deger\,$^{a,b,c,}{\!}$
		\footnote{\tt sadik.deger@boun.edu.tr},
	Marc Geiller\,$^{d,}{\!}$
		\footnote{\tt marc.geiller@ens-lyon.fr},
	Jan Rosseel\,$^{e,c,}{\!}$
		\footnote{\tt rosseelj@gmail.com},
	Henning Samtleben\,$^{d,f,}{\!}$
		\footnote{\tt henning.samtleben@ens-lyon.fr}
		}} \vskip .8cm
	
	{\it ${}^a$ Department of Mathematics, Bogazici University, Bebek, 34342, Istanbul, Turkey}\\[1.5ex] \ 
	{\it ${}^b$ Feza Gursey Center for Physics and Mathematics, Bogazici University, Kandilli, 34684, Istanbul, Turkey}\\[1.5ex] \ 
 {\it ${}^c$ Erwin Schr\"odinger International Institute for Mathematics and Physics, University of Vienna, Boltzmanngasse 9, 1090, Vienna, Austria}\\[1.5ex] \ 
	{\it ${}^d$ Univ Lyon, ENS de Lyon, CNRS, Laboratoire de Physique, F-69342 Lyon, France}\\[1.5ex] \ 
	{\it  $^{e}$ Division of Theoretical Physics, Rudjer Bo\v skovi\'c Institute, Bijeni\v cka 54, 10000 Zagreb, Croatia}\\[1.5ex] \ 
	{\it  $^{f}$ Institut Universitaire de France (IUF)}\\ \ \\
	
\end{center}
%%%%%%%%%%%%%%%%%%%%%%%%%%%%%%%%%%%%%%%%%%%%%%%
\vfill

\begin{center}
	\textbf{Abstract}
	
\end{center}
\begin{quote}
We present an action for minimal massive gravity (MMG) in three dimensions 
in terms of a dreibein and an independent spin connection.
Furthermore, the construction provides an action principle for an infinite family 
of so-called third-way consistent generalizations of the three-dimensional Einstein field equations,
including exotic massive gravity and new higher-order generalizations.
It allows to systematically construct the matter couplings for these models,
including the couplings to fermions, depending on the spin connection.
In particular, we construct different supersymmetric extensions of MMG,
and derive their second order fermionic field equations. This establishes a new class
of three-dimensional supergravity theories and we discuss their limit to topological massive supergravity.
Finally, we identify the landscape of (A)dS vacua of the supersymmetric models.
We analyze the spectrum and the unitarity properties of these vacua.
We recover the known AdS vacua of MMG which are bulk and boundary unitary.
\end{quote}
\vfill
\setcounter{footnote}{0}

%\bigskip
%\bigskip
\end{titlepage}

\tableofcontents \noindent {}
%\newpage

%%%%%%%%%%%%%%%%%%%%%%%%%%%%%%%%%%%%%%%%%%%%%%%%%%%%%%%
%%%%%%%%%%%%%%%%%%%%%%%%%%%%%%%%%%%%%%%%%%%%%%%%%%%%%%%

\section{Introduction} \label{sec:Intro}

This paper gives a detailed account of the results announced in \cite{Deger:2022gim}, on theories of massive gravity in three dimensions. Three-dimensional (3d) gravity, although being topological, has always been a fruitful testbed to investigate various aspects of classical and quantum gravity, in particular in relation to black hole physics and holography \cite{Deser:1983tn,Deser:1983nh,Achucarro:1987vz,Witten:1988hc,MR1637718}. It is for example in the context of 3d AdS spacetimes that Brown and Henneaux have discovered an asymptotic double Virasoro symmetry algebra with non-trivial central charge \cite{Brown:1986nw}, thereby setting the foundations for the AdS/CFT correspondence. The latter has in turn been applied successfully to computations \cite{Carlip:1995cd,Strominger:1997eq,Birmingham:1998jt} of the entropy of 3d BTZ black holes \cite{Banados:1992wn}.

Among the interesting field-theoretic properties of 3d gravity exists the possibility of making the theory massive without breaking diffeomorphism invariance. The simplest embodiment of this mechanism is realized by topologically massive gravity (TMG) \cite{Deser:1982vy,Deser:1981wh}. This is a parity-breaking third order theory obtained by supplementing the Einstein--Hilbert Lagrangian by a Chern--Simons term for the Levi--Civita connection. In \cite{Bergshoeff:2009hq}, this was generalized to a fourth order parity-preserving theory called new massive gravity (NMG), which was in turn further generalized in \cite{Hohm:2012vh} to a model known as general massive gravity (GMG) and which interpolates between TMG and NMG. The fact that NMG propagates two massive gravitons has motivated the authors of \cite{Bergshoeff:2014pca} to search for the most general theory propagating a single graviton only, and this has led to the introduction of minimal massive gravity.

Minimal massive gravity (MMG) is a higher-order generalization of three-dimensional Einstein gravity, described by the field equations~\cite{Bergshoeff:2014pca}
\begin{equation}
\frac1{\mu}\,C_{\mu\nu}
+\bar\sigma\,G_{\mu\nu}
+ \bar\Lambda_0\, g_{\mu\nu}
= 
\frac{\gamma}{2\mu^2} \, \epsilon_{\mu\kappa\lambda}  \epsilon_{\nu\sigma\tau} S^{\kappa\sigma} S^{\lambda\tau}  
\;.
\label{MMGintro}
\end{equation}
Here, $G_{\mu\nu}$, $S_{\mu\nu}$, and  $C_{\mu\nu}$, denote the Einstein tensor,
the Schouten tensor, and the Cotton tensor, respectively, associated with the three-dimensional metric $g_{\mu\nu}$
\begin{equation}
G_{\mu\nu}=R_{\mu\nu}-\tfrac12\,R\,g_{\mu\nu}
\;,\qquad
S_{\mu\nu}= R_{\mu\nu}-\tfrac14\,R\,g_{\mu\nu}
\;,\qquad
C_{\mu\nu}=
\epsilon_{\mu\rho\sigma} \nabla^{\rho} S^{\sigma}{}_{\nu}
\;.
\end{equation}
The coupling constants in (\ref{MMGintro}) are $\{\mu, \bar\sigma, \bar\Lambda_0, \gamma\}$.
In the limit $\gamma\rightarrow0$, the MMG equations reduce to those of topologically massive gravity (TMG) \cite{Deser:1981wh}.
For $\gamma\not=0$, the r.h.s.\ of (\ref{MMGintro}) is an unusual extension of the Einstein field equations. It is on-shell divergence-free, as required for consistency with the l.h.s.. However, unlike in standard gravitational theories this is not a consequence of Bianchi identities or the matter field equations,
but rather follows from iterating the gravitational equations (\ref{MMGintro}) themselves ---
a mechanism dubbed ``third way consistent'' in \cite{Bergshoeff:2014pca} (for a review see \cite{Bergshoeff:2015zga, Deger:2021ojb}).

As a consequence, the MMG equations (\ref{MMGintro}) cannot be derived from a standard action principle of the metric (or the dreibein) alone. 
An action principle based on a first order Lagrangian with auxiliary fields has been given in \cite{Bergshoeff:2014pca}, in the region of
parameter space where
\begin{equation}
\mu^2\,(1+\gamma\bar\sigma)^2 > \gamma^3\,\bar\Lambda_0
\;,
\label{ktnintro}
\end{equation}
see (\ref{ktn}) below.
Absence of a standard action functional in particular obscures the coupling of MMG to matter.
In particular, the standard matter energy-momentum tensor does not yield a consistent source for equations (\ref{MMGintro}),
rather the presence of matter modifies these equations by a source tensor quadratic in the energy-momentum 
tensor~\cite{Arvanitakis:2014yja,Ozkan:2018cxj}. 

In reference \cite{Ozkan:2018cxj}, equations (\ref{MMGintro}) were generalized to an infinite sequence of third-way consistent gravitational 
models with the next-to-simplest example dubbed Exotic Massive Gravity (EMG).
Third way consistent deformations also exist for gauge theories, they were constructed in \cite{Arvanitakis:2015oga}
for three-dimensional Yang-Mills theory, and in \cite{Broccoli:2021pvv} for higher-dimensional $p$-form theories.
The minimal supersymmetric extension of the three-dimensional gauge theory, and its coupling to supergravity were obtained in \cite{Deger:2021fvv,Deger:2022znj}.

In the present paper, we construct a new and universal action principle for third-way consistent gravitational models, 
including minimal massive gravity and the higher order generalizations thereof. 
This allows to streamline and systematically construct the coupling of these models to bosonic and fermionic matter.
Starting from this action, we construct the supersymmetric extension of MMG: minimal massive supergravity (MMSG). We perform this construction up to and including quartic order in the fermions.
This is the first example of a supersymmetric third-way consistent gravitational model and
constitutes a new class of three-dimensional supergravities, which can be seen as deformations of topologically massive supergravity (TMSG) \cite{Deser:1982sw,Deser:1984cts}.
We study its maximally symmetric vacua of (A)dS and Minkowski type, and in particular the interplay of supersymmetry and unitarity. 
A preliminary version of these results was announced in \cite{Deger:2022gim}.

The rest of this paper is organized as follows. In section~\ref{sec:bosonic} we present a universal action principle for the third-way consistent gravitational models in terms of a dreibein and an independent spin connection. The action allows to systematically and algorithmically produce models of three-dimensional massive gravity of increasingly higher order. We show how to recover in this framework the original minimal massive gravity (\ref{MMGintro}) and reproduce the action of \cite{Bergshoeff:2014pca}. The construction also provides actions for the models of \cite{Ozkan:2018cxj} and generalizations thereof, without increasing the number of fields in the Lagrangian. We discuss the general matter couplings of these models, including the couplings to fermions, which may in general depend on the Levi-Civita as well as on the independent spin connection. We show explicitly how Lorentz and diffeomorphism symmetry of the Lagrangian guarantee consistency of the resulting field equations.
In section~\ref{sec:susy}, we focus on the MMG model and construct its minimal ${\cal N}=(1,0)$ supersymmetric extension (MMSG). 
The fermionic sector of the model carries two gravitino fields, sharing one local supersymmetry. Reminiscent of the first order formulation of TMSG \cite{Sezgin:2009dj,Routh:2013uc},
the first order fermionic Lagrangian then propagates a massive spin-3/2 mode as a superpartner to the massive spin-2 mode.
Upon elimination of one of the gravitino fields, we obtain the second-order fermionic field equations, which constitute the ``superpartner'' to the bosonic  equations (\ref{MMGintro}). We discuss the limit in which MMSG reduces to topologically massive gravity (TMSG) \cite{Deser:1982sw,Deser:1984cts}, 
both on the level of the action and the field equations.

In section~\ref{sec:AdSvacua}, we study the parameter space of supersymmetric MMSG models and the landscape of their (A)dS vacua. We show that every bosonic MMG model that possesses an AdS vacuum admits up to four supersymmetric extensions, parametrized by the real roots of a quartic equation (\ref{lambda_eta}). A given AdS vacuum can be supersymmetric in one supersymmetric extension, and non-supersymmetric in another. We compute the central charges as well as the bosonic and fermionic mass spectra around all AdS vacua. In particular, the analysis recovers the AdS vacua of \cite{Bergshoeff:2014pca} that are both bulk and boundary unitary. In our previous construction \cite{Deger:2022gim}, all supersymmetries were broken around the unitarity vacua. The present analysis however reveals a larger parameter space of supersymmetric models, and in particular an additional region of parameters in which the unitary vacua do preserve half of the supersymmetry.

Our conventions and notations are as follows. We denote the 3-dimensional spacetime indices by $\mu,\nu,\dots$, and local Lorentz indices by $a,b,\dots$. Our choice of signature is $(-,+,+)$. The Levi-Civita tensor density and tensor are denoted by $\varepsilon^{\mu\nu\rho}$ and $\epsilon^{\mu\nu\rho}$ respectively. We denote anti-symmetrization by $A_{[a}B_{b]}=(A_aB_b-A_bB_a)/2$.

%%%%%%%%%%%%%%%%%%%%%%%%%%%%%%%%%%%%%%%%%%%%%%%%%%%%%%%
%%%%%%%%%%%%%%%%%%%%%%%%%%%%%%%%%%%%%%%%%%%%%%%%%%%%%%%

\section{Bosonic Lagrangians}
\label{sec:bosonic}

%%%%%%%%%%%%%%%%%%%%%%%%%%%%%%%%%%%%%%%%%%%%%%%%%%%%%%%

In this first section we present the bosonic sector of the theory. Building up on \cite{Bergshoeff:2014pca,Ozkan:2018cxj}, we introduce a Lagrangian formalism which enables to produce so-called third-way consistent theories through the iteration of their equations of motion. We then explain how this mechanism allows to recover MMG in a particular case, and also provide a first order formulation using two independent spin connections. This latter is the key to the supersymmetrization of the model. We also explain how our general mechanism can be used to systematically and algorithmically produce models of massive gravity which are increasingly higher order, in a way similar (yet more general) to what was outlined in \cite{Ozkan:2018cxj}. We also give explicit new examples. Finally, we discuss the coupling to bosonic and fermionic matter and determine conditions on the matter current for the third way consistency to work when matter Lagrangian also depends on the spin connection .

\subsection{General mechanism for third way consistent field equations}
\label{subsec:general}

As our starting point, let us consider a dreibein $e_\mu{}^{a}$ and an independent (and thus torsionful) spin connection $\varpi_\mu{}^{a}$. This field content is then used to build the general class of 3-dimensional bosonic Lagrangians
\begin{align}\label{LMasterEW}
 {\cal L}[e,\varpi]=\,
{\cal L}_{0}[e]+ \tau\,\varepsilon^{\mu\nu\rho}\,
    e_\mu{}^{a}{} D[\varpi]_\nu  e_{\rho a} 
+\kappa\,{\cal L}_{\rm CS}[\varpi] 
     \;,
\end{align}
where ${\cal L}_{0}[e]$ is at this stage an arbitrary gravitational Lagrangian depending only on the dreibein $e$. The term with coupling constant $\tau$ is a torsion term, while that with coupling $\kappa$ is the ${\rm SO}(2,1)$ Chern-Simons Lagrangian 
\begin{equation}
{\cal L}_{\rm CS}[\varpi] = 
\varepsilon^{\mu\nu\rho}
\left(
\varpi_\mu{}^{a} \partial_\nu \varpi_{\rho a}
+\frac{1}{3}\varepsilon_{a b c}\, \varpi_\mu{}^{a} \varpi_\nu{}^{b} \varpi_\rho{}^{c}
\right)
,
\label{LCS}
\end{equation}
for the connection $\varpi$. The covariant derivative with respect to $\varpi$ is denoted by $D[\varpi]_\mu$, while its torsion and curvature are defined as
\begin{subequations}
\begin{align}
T[\varpi]_{\mu\nu}{}^a \coloneqq&\, 2\, D[\varpi]_{[\mu} e_{\nu]}{}^a
=2\,\partial _{[\mu} e_{\nu]}{}^a + 2\,{\varepsilon^a}_{bc} \,\varpi_{[\mu}{}^{b} e_{\nu]}{}^{c}\label{def:torsion}
\;,\\
R[\varpi]_{\mu\nu}{}^a \coloneqq&\, 2\partial_{[\mu} \varpi_{\nu]}{}^a + {\varepsilon^a}_{b c} \varpi_\mu{}^b \varpi_\nu{}^c
\;,
\end{align}
\end{subequations}
respectively. The Lagrangian (\ref{LMasterEW}) is manifestly invariant under 
Lorentz transformations
\bea
\delta_\Lambda e_\mu{}^{a} = 
\varepsilon^{abc}\,e_{\mu b}\, \Lambda_{c}
\;,\quad
\delta_\Lambda \varpi_\mu{}^{a}  &=&
D[\varpi]_\mu \Lambda^{a}
\;.
\label{Lorentz}
\eea

We can now study the equations of motion of the Lagrangian (\ref{LMasterEW}). Taking variations with respect to the connection $\varpi$ yields the curvature equation
\begin{equation}
R[\varpi]_{\mu\nu}{}^a +\frac{\tau}{\kappa}\,{\varepsilon^a}_{b c}\,
    e_\mu{}^{b}{}   e_{\nu}{}^{c} =0
    \;.
    \label{RbEE}
\end{equation}
The equation of motion imposed by the dreibein, on the other hand, is the torsion equation
\begin{equation}
2\,\mathbb{G}^{\mu a}
+ \tau\,\epsilon^{\mu\nu\rho}\,T[\varpi]_{\nu\rho}{}^a=0
\;,
\label{TG}
\end{equation}
where $\mathbb{G}^\mu{}_{a}$ is defined by
\begin{equation}
\delta{\cal L}_0[e] = 2\,\sqrt{-g}\,\mathbb{G}^\mu{}_{a}\,\delta e_\mu{}^{a}
\;.
\label{delL}
\end{equation}
Diffeomorphism and Lorentz symmetry imply that the tensor 
$\mathbb{G}_{\mu\nu}\coloneqq \mathbb{G}_{\mu}{}^{a}e_{a \nu}$ 
is divergence-free and symmetric, i.e. that we have
\begin{equation}
\mathbb{G}_{\mu\nu}=\mathbb{G}_{\nu\mu}\;,\qquad
\nabla^\mu\mathbb{G}_{\mu\nu} = 0
\;.
\label{ddG}
\end{equation}
In order to combine the equations of motion \eqref{RbEE} and \eqref{TG} into a single metric field equation, we now decompose the connection $\varpi$ in terms of the torsionless Levi-Civita connection $\mathring{\omega}$ and the contorsion tensor defined as
\begin{align}
  K[\varpi]_{\mu}{}^{a}\coloneqq\varpi_\mu{}^{a}-\mathring{\omega}_\mu{}^{a}\;.
\end{align}
%\begin{align}
%  K[\varpi]_{\mu}{}^{a} 
%=\;&
%\varpi_\mu{}^{a}-\omega_\mu{}^{a}
%\nonumber\\
%=\;&
%\tfrac12\,\epsilon^{\rho\sigma\tau} \left(
%e_\mu{}^{b}e_\rho{}^{a}  -\tfrac12 e_\mu{}^{a}  e_\rho{}^{b}\right)
%T[\varpi]_{\sigma\tau, b}
%\;,
%\end{align}
Since $\mathring{\omega}$ is compatible with $e$, we get from the definition \eqref{def:torsion} of the torsion that
\be
T[\varpi]_{\mu\nu}{}^a =2\,{\varepsilon^a}_{bc} \,K[\varpi]_{[\mu}{}^{b} e_{\nu]}{}^{c}.
\ee
Inverting this relation, we can then rewrite the field equations (\ref{TG}) as
\begin{equation}
K[\varpi]_{\mu}{}^{a} = -\frac1{\tau}\left(\mathbb{G}_{\mu}{}^{a} - \frac12e_{\mu}{}^{a}\,\mathbb{G} \right)
\eqqcolon -\frac1{\tau}\, \mathbb{S}_{\mu}{}^{a}
\;,
\label{KS}
\end{equation}
where $\mathbb{G}\coloneqq \mathbb{G}_\mu{}^{a}e_{a}{}^\mu$\,. Finally, using the relation between the contorsion and the curvatures of $\varpi$ and $\mathring{\omega}$, i.e.
\begin{equation}
R[\varpi]_{\mu\nu}{}^{a}
= R[\mathring{\omega}]_{\mu\nu}{}^{a} +2\,D[\mathring{\omega}]_{[\mu}   K[\varpi]_{\nu]}{}^{a}
+\varepsilon^{a b c}\,K[\varpi]_{\mu b} K[\varpi]_{\nu c}
\;,
\label{RRKK}
\end{equation}
we can combine (\ref{RbEE}) and (\ref{KS}) to obtain the general field equations
\begin{equation}
G_{\mu\nu}
-\frac{\tau}{\kappa}\,g_{\mu\nu}
=
\frac{1}{\tau}\mathbb{C}_{\mu\nu} -\frac1{2\tau^2}\,\epsilon_{\mu\kappa\rho}  \epsilon_{\nu\lambda\sigma}\,\mathbb{S}^{\kappa\lambda}\mathbb{S}^{\rho\sigma}\,,
\label{eom_final}
\end{equation}
where $G_{\mu\nu}$ is the Einstein tensor, $\mathbb{S}_{\mu\nu}={\mathbb{S}_\mu}^ae_{\nu a}$ is defined by (\ref{KS}) and (\ref{delL}), and we have introduced 
\begin{equation}
\mathbb{C}_{\mu\nu}\coloneqq\epsilon_{\mu\rho\sigma} \,\nabla^{\rho}   \mathbb{S}^{\sigma}{}_\nu
\;.
\label{defC}
\end{equation} 
Symmetry of $\mathbb{C}_{\mu\nu}$ is a consequence of (\ref{ddG}), which implies
\be\label{S property}
\nabla^\mu\mathbb{S}_{\mu\nu}=\nabla_\nu\mathbb{S}\;,
\quad
\mbox{with}\;\; \mathbb{S}=g^{\mu\nu}\,\mathbb{S}_{\mu\nu}
\;.
\ee
The field equations (\ref{eom_final}) depend only on the metric.
Importantly, these field equations in purely metric form cannot be obtained from the variational principle of a Lagrangian in metric variables. Relatedly, we should in particular note that while the left-hand side is divergence-free by virtue of the Bianchi identities, this is not manifestly the case for the right-hand side. Instead, the consistency requirement of vanishing divergence of the right-hand side requires to iterate \eqref{eom_final} itself. From the definition of $\mathbb{C}_{\mu\nu}$ we get $\nabla^\mu\mathbb{C}_{\mu\nu}=\epsilon_{\nu\rho\sigma}R^{\rho\tau}{\mathbb{S}_\tau}^\sigma$. With this and (\ref{S property}), up to a rescaling by $\tau$, 
the divergence of the right-hand side of \eqref{eom_final} becomes
\be\label{consistency}
\nabla^\mu\mathbb{C}_{\mu\nu} -\frac1{2\tau}\,\epsilon_{\mu\kappa\rho}  \epsilon_{\nu\lambda\sigma}\,\nabla^\mu(\mathbb{S}^{\kappa\lambda}\mathbb{S}^{\rho\sigma})\,=\epsilon_{\nu\rho\sigma}R^{\rho\kappa}{\mathbb{S}_\kappa}^\sigma-\frac{1}{\tau}\epsilon_{\nu\kappa\sigma}{\mathbb{C}_\rho}^\kappa\mathbb{S}^{\rho\sigma}\stackrel{\eqref{eom_final}}{=}0\;,
\ee
where crucially in the last step the field equations \eqref{eom_final} must be used to replace 
$\mathbb{C}_\rho^{\, \, \kappa}$ in order to obtain a vanishing result. This consistency mechanism was coined ``third-way consistency'' in \cite{Bergshoeff:2014pca}. Note that $\kappa \tau \neq 0$ is needed for this procedure to work.

A key feature of the variational principle outlined above leading to \eqref{eom_final} is that it requires the use of the integrability conditions (\ref{RRKK}). A similar mechanism has been employed in \cite{Deger:2022znj} in order to reformulate the third-way consistent deformation of Yang-Mills theory \cite{Arvanitakis:2015oga} in terms of a gauged scalar sigma model. 
{}From this systematic analysis we can now recover known models such as MMG but also higher order generalizations.

\subsection{Minimal massive gravity}

We now focus on recovering MMG \cite{Bergshoeff:2014pca}. From the general form of the field equations \eqref{eom_final}, one can already anticipate that MMG is obtained in the case where $\mathbb{S}_{\mu\nu}$ is related to the Schouten tensor and $\mathbb{C}_{\mu\nu}$ to the Cotton tensor. This can be achieved by choosing in \eqref{LMasterEW} the Lagrangian
\begin{equation}
{\cal L}_0[e] =
\varepsilon^{\mu\nu\rho}\left(\frac1{G_3}\,e_\mu{}^{a}R[\mathring\omega]_{\nu\rho, a}
+\lambda\,\varepsilon_{a b c} e_\mu{}^{a} e_\nu{}^{b} e_\rho{}^{c}\right)
\;,
\label{L0MMG}
\end{equation}
where as before, $\mathring\omega$ is the torsionless Levi-Civita connection determined by $e_\mu{}^a$.
For simplicity we will now set the gravitational constant to $G_3=1$. 
The full Lagrangian (\ref{LMasterEW}) thus takes the form
\begin{equation}
{\cal L}[e,\varpi] =
\varepsilon^{\mu\nu\rho}\Big(e_\mu{}^{a}R[\mathring{\omega}]_{\nu\rho, a}
+\lambda\, \varepsilon_{a b c} e_\mu{}^{a} e_\nu{}^{b} e_\rho{}^{c}
+ \tau\,e_\mu{}^{a}{} D[\varpi]_\nu  e_{\rho a} \Big)
+\kappa\,{\cal L}_{\rm CS}[\varpi] 
\;.
\label{LMMG}
\end{equation}
For later use, let us explicitly spell out the equations of motion obtained by variation as
\bea
\delta {\cal L}&=&
\varepsilon^{\mu\nu\rho}\,\delta e_\mu{}^{a}\,{\cal E}_{\nu\rho,a}
+\kappa\,\varepsilon^{\mu\nu\rho}\,\delta \varpi_\mu{}^{a} \bar{\cal E}_{\nu\rho,a}
\;,
\eea
with
\bea
{\cal E}_{\mu\nu,a} &=& 
R[\mathring{\omega}]_{\mu\nu,a}
+ \tau\,T[\varpi]_{\mu\nu,a}
+3\,\lambda\,\varepsilon_{abc}\,e_\mu{}^{b} e_\nu{}^{c}
\;,\nonumber\\[1ex]
\bar{\cal E}_{\mu\nu,a} &=& 
R[\varpi]_{\mu\nu,a} 
+ \frac{\tau}{\kappa}\,\varepsilon_{abc}\,
    e_\mu{}^{b}{}   e_{\nu}{}^{c}   \;.
    \label{eom_bosonic}
\eea
Let us also note that the Lagrangian (\ref{LMMG}) scales homogeneously under
the transformation
\bea
&&e_\mu{}^{a} \rightarrow \sigma e_\mu{}^{a}\;,\quad
\varpi_\mu{}^{a} \rightarrow \varpi_\mu{}^{a}\;,\quad
\nonumber\\
&&\lambda \rightarrow \sigma^{-2} \lambda\;,\quad
\tau \rightarrow \sigma^{-1} \tau\;,\quad
\kappa \rightarrow \sigma \kappa\;,
\label{scaling}
\eea
of fields and coupling constants, where $\sigma$ is a nonzero constant.

{}From \eqref{TG}, \eqref{KS}, \eqref{defC} we obtain that\be\label{MMG S tensor}
\mathbb{S}_{\mu\nu}=\frac{3\lambda}{2}g_{\mu\nu}+S_{\mu\nu}\;,
\quad\quad
S_{\mu\nu}\coloneqq R_{\mu\nu}-\frac14R\,g_{\mu\nu}\;,
\quad\quad
\mathbb{C}_{\mu\nu}=C_{\mu\nu}\coloneqq
\epsilon_{\mu\rho\sigma} \nabla^{\rho} S^{\sigma}{}_{\nu}\;,
\ee
where $S_{\mu\nu}$ and $C_{\mu\nu}$ are the Schouten and Cotton tensor, respectively. 
The field equations \eqref{eom_final} then specialize to the MMG equations
\be
\left(1+\frac{3\lambda}{2\tau^2}\right)G_{\mu\nu}-\left(\frac{\tau}{\kappa}+\frac{9\lambda^2}{4\tau^2}\right)g_{\mu\nu}=\frac{1}{\tau}C_{\mu\nu}+\frac{1}{\tau^2} J_{\mu\nu}\,,
\label{MMGh}
\ee
where following \cite{Bergshoeff:2014pca,Ozkan:2018cxj} we have denoted
\be
J_{\mu\nu}\coloneqq-\frac{1}{2}\epsilon_{\mu\kappa\lambda}  \epsilon_{\nu\sigma\tau} S^{\kappa\sigma} S^{\lambda\tau}\,.
\ee
Performing the redefinitions
\be\label{param}
\left(1+\frac{3\lambda}{2\tau^2}\right)=-\frac{\mu\bar{\sigma}}{\tau}\,,
\quad\quad
\left(\frac{\tau}{\kappa}+\frac{9\lambda^2}{4\tau^2}\right)=\frac{\mu\bar{\Lambda}_0}{\tau}\,,
\quad\quad
\frac{1}{\tau}=\frac{\gamma}{\mu}\,,
\ee
one obtains
\begin{equation}
\bar\sigma\,G_{\mu\nu}
+ \bar\Lambda_0\, g_{\mu\nu}
= 
-\frac1{\mu}\,C_{\mu\nu}-\frac{\gamma}{\mu^2} J_{\mu\nu}\;,
\label{MMG0}
\end{equation}
which is the MMG field equation (1.5) of \cite{Bergshoeff:2014pca}. Taking the limit $\gamma\rightarrow0$, we obtain from (\ref{MMG0}) the equations of motion of TMG, which will therefore be included in our supersymmetrization as well. We will see however that in the fermionic sector the limit requires a more careful discussion.

Let us now turn to the first order formulation of the model, which is obtained by replacing (\ref{L0MMG}) by the Palatini Lagrangian
\begin{equation}
{\cal L}_0[e,\omega] =
\varepsilon^{\mu\nu\rho}\Big(e_\mu{}^{a}R[\omega]_{\nu\rho, a}
+\lambda \,\varepsilon_{a b c} e_\mu{}^{a} e_\nu{}^{b} e_\rho{}^{c}\Big)\,,
\label{L0MMG_Pal}
\end{equation}
where now $\omega$ is an independent connection. Upon solving the field equations for $\omega$ one finds $\omega=\mathring\omega$, showing the equivalence to (\ref{L0MMG}). With the choice (\ref{L0MMG_Pal}), the total Lagrangian (\ref{LMasterEW}) is the sum of the so-called ``standard'' and the ``exotic'' actions for 3-dimensional gravity \cite{Witten:1988hc}, however now with both sectors carrying different spin connections $\omega$, and $\varpi$, respectively. In this first order formulation of MMG, it is precisely the use of two independent spin connections which is responsible for the appearance of the massive degree of freedom. Indeed, for $\omega=\varpi$ the Lagrangian (\ref{LMasterEW}) corresponds simply to a reformulation of 3-dimensional (topological) gravity, such as studied in \cite{Mielke:1991nn,Cacciatori:2005wz,Geiller:2020edh}. In the formulation with two independent connections, when $\kappa\tau<0$ we can consider the field redefinition
\begin{equation}
\omega_\mu{}^{a} = \Omega_\mu{}^{a}+ \alpha\,h_\mu{}^{a}    
\;,
\quad\quad
\varpi_\mu{}^{a} = \Omega_\mu{}^{a} +\theta\,e_\mu{}^{a}
\;,
\label{oOmega}
\end{equation}
in terms of two new fields $\Omega_\mu{}^a$ and $h_\mu{}^a$. With this, the Lagrangian (\ref{LMasterEW}) 
takes the form
\bea
{\cal L}[e,\Omega,h]&=&
(1+\kappa\,\theta)\,\varepsilon^{\mu\nu\rho}\,e_\mu{}^{ a}\,R[\Omega]_{\nu\rho, a}
+(\tau+{\kappa}\,\theta^2)\,\varepsilon^{\mu\nu\rho}\,e_\mu{}^{ a}\,D[\Omega]_{\nu} e_{\rho, a}
\nonumber\\
&&{}
+2\,\alpha\,\varepsilon^{\mu\nu\rho}\,{h}_{\mu}{}^{ a}\,D[\Omega]_\nu e_{\rho  a}
+\alpha^2\,\varepsilon^{\mu\nu\rho}\varepsilon_{ a b c} \,e_\mu{}^{ a}\,{h}_{\nu}{}^{ b} {h}_{\rho}{}^{ c}
\nonumber\\
&&{}
+\left(\lambda+ \tau\,\theta+\frac13{\kappa\,\theta^3}\right)
\,\varepsilon^{\mu\nu\rho}\,\varepsilon_{ a b c}\,e_\mu{}^{ a}e_\nu{}^{ b}e_\rho{}^{ c}
+{\kappa}\,{\cal L}_{\rm CS}[\Omega]
\;.
\label{LMMGV3}
\eea
Choosing $\theta=\frac{\sqrt{-\kappa\tau}}{\kappa}$, this precisely reproduces the first-order Lagrangian of \cite{Bergshoeff:2014pca} and we read off the identification with the parameters 
of~\cite{Bergshoeff:2014pca} as\footnote{Recall that $\sigma^2=1$ in \cite{Bergshoeff:2014pca}.}
\bea \label{identication}
\mu = \frac{\alpha}{\kappa}\;,\quad
\alpha\,\sigma = -1\mp\sqrt{-\kappa\tau}\;,\quad
\alpha\,\Lambda_0 = 3\,\lambda\pm \frac{2\,\tau}{\kappa}\,\sqrt{-\kappa\tau}
\;.
\eea
In particular, this shows that the action of~\cite{Bergshoeff:2014pca} only covers the 
parameter space region $\kappa\tau<0$\,. 
In terms of the original parameters appearing in \eqref{MMG0}, the condition $\kappa\tau<0$ translates into the condition 
\begin{equation}
\mu^2\,(1+\gamma\bar\sigma)^2 > \gamma^3\,\bar\Lambda_0
\;.
\label{ktn}
\end{equation}
On the other hand, the action (\ref{LMMG}) describes the MMG equations \eqref{MMG0} for any value of its parameters
since (\ref{param}) always admits a solution for $\{\kappa, \tau, \lambda\}$.

We have now at our disposal all the ingredients to describe the bosonic sector of our theory. In particular, the supersymmetric extension which we consider in the next section will be built from \eqref{LMasterEW} and \eqref{L0MMG_Pal}, with a connection $\omega$ determined by the torsion equation in terms of the fermions. Before moving to the supersymmetric Lagrangian, let us close this section with a discussion of the extension to higher order models of massive gravity, as well as the matter couplings.

\subsection{Higher order massive gravities}

Tracing back the argument leading to the consistency condition \eqref{consistency} for the field equations, one can see that it follows essentially from the property \eqref{S property}, which itself comes from \eqref{ddG} and the definition \eqref{KS}. Crucially, since \eqref{S property} is linear in the tensor $\mathbb{S}_{\mu\nu}$, the consistency of the field equations is therefore preserved if one considers any linear combination of such tensors. This enables to describe via this mechanism all the known third way consistent models of massive gravity, and also to straightforwardly produce higher order extensions.

For example, a family of known models can be recovered if we use couplings $\alpha_i$ for the metric, the Schouten, and the Cotton tensors, and consider
\be\label{S of EMGMG}
\mathbb{S}_{\mu\nu}=\alpha_1\,g_{\mu\nu}+\alpha_2\,S_{\mu\nu}+\alpha_3\,C_{\mu\nu}\,.
\ee
Taking $\alpha_2=\alpha_3=0$ leads to the Einstein equations with cosmological constant. Taking $\alpha_3=0$ as in \eqref{MMG S tensor} leads to MMG as discussed in the previous section. Taking $\alpha_1=\alpha_2=0$ leads to the so-called `Exotic Massive Gravity' (EMG), while taking only $\alpha_2=0$ leads to its parity-violating generalization dubbed `Exotic General Massive Gravity' (EGMG) \cite{Ozkan:2018cxj}. The linearity argument given above shows that even with all three arbitrary couplings one obtains consistent field equations which are of $4^\text{th}$ order. Explicitly, using \eqref{S of EMGMG} in \eqref{eom_final} produces the field equations
\begin{align}\label{EOM of EMGMG}
\left(1+\frac{\alpha_1\alpha_2}{\tau^2}\right)G_{\mu\nu}-\left(\frac{\tau}{\kappa}+\frac{\alpha_1^2}{\tau^2}\right)\,g_{\mu\nu}
&=\frac{1}{\tau}\left(\alpha_2-\frac{\alpha_1\alpha_3}{\tau}\right)C_{\mu\nu}+\frac{\alpha_3}{\tau}H_{\mu\nu}-\frac{\alpha_3^2}{\tau^2}L_{\mu\nu}\quad\cr
&\;\;\;+\frac{\alpha_2^2}{\tau^2}J_{\mu\nu}-\frac{\alpha_2\alpha_3}{\tau^2}\,\epsilon_{\mu\kappa\rho}  \epsilon_{\nu\lambda\sigma}\,S^{\kappa\lambda}C^{\rho\sigma}\,,
\end{align}
where following the notations of \cite{Ozkan:2018cxj} we have further introduced
\be
H_{\mu\nu}\coloneqq\epsilon_{\mu\rho\sigma} \,\nabla^{\rho}C^{\sigma}{}_\nu\,,
\quad\quad\quad
L_{\mu\nu}\coloneqq\frac{1}{2}\,\epsilon_{\mu\kappa\rho}  \epsilon_{\nu\lambda\sigma}C^{\kappa\lambda}C^{\rho\sigma}\,.
\ee
They are third way consistent by the mechanism described in section~\ref{subsec:general}.
Upon redefinition of the constants,
$\alpha_1=\frac{\tau m^2}{\mu}$,
$\alpha_2=\tau\tilde{\alpha}$,
$\alpha_3=\frac{\tau}{m^2}$,
 $\Lambda=-\left(\frac{\tau}{\kappa}+\frac{\alpha_1^2}{\tau^2}\right)$,
 equations (\ref{EOM of EMGMG}) take the form
\begin{equation}\label{EOM of EMGMG new}
\Lambda\,g_{\mu\nu}+
\left(1+\tfrac{m^2\tilde{\alpha}}{\mu}\right)G_{\mu\nu}
+\left(\tfrac{1}{\mu}-\tilde{\alpha}\right)C_{\mu\nu}-\frac{1}{m^2}H_{\mu\nu}+\frac{1}{m^4}L_{\mu\nu}
=\tilde{\alpha}^2 J_{\mu\nu}-\tfrac{\tilde{\alpha}}{m^2}\,\epsilon_{\mu\kappa\rho}  \epsilon_{\nu\lambda\sigma}\,S^{\kappa\lambda}C^{\rho\sigma}\,.
\end{equation}
The l.h.s.\ of \eqref{EOM of EMGMG new} for $\tilde\alpha=0$, precisely reproduces the field equations of EGMG appearing in equation (1.8) of \cite{Ozkan:2018cxj}. A non-vanishing $\tilde\alpha$ on the other hand describes a further one-parameter deformation of EGMG which we might dub `Exotic More General Massive Gravity' (EMGMG). 
As described above, the field equations \eqref{EOM of EMGMG new} can be obtained from our variational principle \eqref{LMasterEW}. 
The Lagrangian $\mathcal{L}_0[e]$ leading to \eqref{eom_final} with \eqref{S of EMGMG} is now itself third order and given by
\be\label{EMGMG Lagrangian}
\mathcal{L}_0[e]=\varepsilon^{\mu\nu\rho}\left(\alpha_2\,e_\mu{}^{a}R[e]_{\nu\rho, a}+\frac{2\alpha_1}{3}\,\varepsilon_{a b c} e_\mu{}^{a} e_\nu{}^{b} e_\rho{}^{c}+\frac{\alpha_3}{2}\,\left[\Gamma^\lambda_{\rho\sigma}\partial_\mu\Gamma^\sigma_{\nu\lambda}+\frac{2}{3}\,\Gamma^\lambda_{\rho\sigma}\Gamma^\sigma_{\mu\tau}\Gamma^\tau_{\nu\lambda}\right]\right),
\ee
where the last term is a Chern-Simons term for the Levi-Civita connection $\Gamma[g]$, as in TMG, whose dependency on the dreibein must be understood through $g_{\mu\nu}={e_\mu}^ae_{\nu a}$.

One can extend the construction to $5^\text{th}$ and higher order field equations by using higher order tensors in \eqref{S of EMGMG}, or equivalently starting from a Lagrangian $\mathcal{L}_0[e]$ including higher order scalars involving the Ricci tensor and the metric. 
In general, including tensors of $n^\text{th}$ order in ${\cal L}_0$ (thus in \eqref{S of EMGMG}), 
the 3rd way consistent theory (\ref{eom_final}) will be of order $n+1$.
For example, we may consider
\be\label{4th order L0}
\mathcal{L}_0[e]=\mathcal{L}_0[e]\big|_{\eqref{EMGMG Lagrangian}}+\sqrt{-g}\,\big(\alpha_4\,R^2+\alpha_5\,R_{\mu\nu}R^{\mu\nu}\big)\;.
\ee
In the particular case where $\alpha_4=-3\alpha_5/8$, the second Lagrangian density on the right-hand side corresponds to that of New Massive Gravity (NMG) \cite{Bergshoeff:2009hq}. One can of course also consider $\alpha_4$ and $\alpha_5$ as independent, compute their contribution to $\mathbb{G}_{\mu\nu}$, then to $\mathbb{S}_{\mu\nu}$, and finally to the general field equations \eqref{eom_final}.
In doing so, from \eqref{4th order L0} we obtain
\begin{align}
\mathbb{G}_{\mu\nu}
&=-2\alpha_1\,g_{\mu\nu}+\alpha_2\,G_{\mu\nu}+\alpha_3\,C_{\mu\nu}\\
&\;\;\;\;+\alpha_4\left(2RR_{\mu\nu}+2g_{\mu\nu}\Box R-2\nabla_\mu\nabla_\nu R-\frac{1}{2}R^2g_{\mu\nu}\right)\nonumber\\
&\;\;\;\;+\alpha_5\left(\frac{3}{2}(R_{\kappa\lambda}R^{\kappa\lambda})g_{\mu\nu}-4R_{\mu\rho}{R^\rho}_\nu-R^2g_{\mu\nu}+\Box R_{\mu\nu}+\frac{1}{2}g_{\mu\nu}\Box R-\nabla_\mu\nabla_\nu R+3RR_{\mu\nu}\right),\nonumber
\end{align}
and when $\alpha_4=-3\alpha_5/8$ this simplifies to
\be
\mathbb{G}_{\mu\nu}=-2\alpha_1\,g_{\mu\nu}+\alpha_2\,G_{\mu\nu}+\alpha_3\,C_{\mu\nu}+\frac{\alpha_5}{2}\,K_{\mu\nu}\,,
\ee
with $K_{\mu\nu}$ given by equation (4) of \cite{Bergshoeff:2009hq} and enjoying the property $g^{\mu\nu}K_{\mu\nu}=K=G^{\mu\nu}S_{\mu\nu}$.

Here, we will not study further the field-theoretical properties of the general field equations \eqref{EOM of EMGMG new}, nor of that arising from the higher order extensions such as \eqref{4th order L0}, but it would be interesting to come back to this in future work. In particular, one should investigate the propagating modes by linearizing the field equations around Minkowski, and study the possible unitarity properties. A $6^\text{th}$ order exotic gravity theory was constructed in \cite{Afshar:2019npk} and should also find its place in this general pattern.

\subsection{Matter couplings}
\label{subsec:matter}

Matter contributions can easily be incorporated in the variational principle arising from \eqref{LMasterEW}, and still lead to third way consistent field equations. In order to see this, let us consider the matter coupling obtained by
\begin{align}\label{matter coupling}
{\cal L}[e,\varpi] & \longrightarrow\; {\cal L}[e, \varpi] + {\cal L}_{\rm matter}[e, \varpi, \Phi]\;,
\end{align}
where $ {\cal L}_{\rm matter}[e, \varpi, \Phi]$ denotes the matter Lagrangian for arbitrary bosonic or fermionic matter fields collectively denoted by $\Phi$. Note that only fermionic matter can couple covariantly to the spin connection $\varpi_\mu{}^a$.
In order to work out its effect on the field equations, we introduce
\begin{equation}
\delta {\cal L}_{\rm matter} = 2\,\sqrt{-g}\left(
\delta e_\mu{}^a T^\mu{}_a +  \delta \varpi_\mu{}^a U^\mu{}_a
\right)
.
\label{TU}
\end{equation}
In the case of bosonic matter only we have $U^\mu{}_a=0$, and the standard energy-momentum tensor $T_{\mu\nu}$ 
is symmetric and covariantly-conserved on-shell by virtue of the Noether identity arising from diffeomorphism invariance of the Lagrangian.
Fermionic matter on the other hand can couple to the spin connection $\varpi_\mu{}^a$ and we will see below that supersymmetry indeed
requires the presence of such couplings. In that case, $T_{\mu\nu}$ is no longer symmetric or conserved, yet the final field equations are (third way) consistent
as we shall verify here.

Replacing (\ref{LMasterEW}) by (\ref{matter coupling}) and repeating the construction described in section~\ref{subsec:general}, it is straightforward to derive that the resulting field equations (\ref{eom_final}) get modified into
\begin{equation} 
G_{\mu\nu}
-\frac{\tau}{\kappa}\,g_{\mu\nu}
- \frac1\tau \mathbb{C}_{\mu\nu} 
+\frac1{2\tau^2}\, \epsilon_{\mu\sigma\tau} \epsilon_{\nu\kappa\lambda}\,\mathbb{S}^{\kappa\sigma} \mathbb{S}^{\lambda\tau} 
=
\mathbb{T}_{\mu\nu} \;,
\label{eom_T}
\end{equation}
with the source tensor $\mathbb{T}_{\mu\nu}$ given by
\begin{equation} 
\mathbb{T}_{\mu\nu} =
\frac1\tau \epsilon_{\mu\sigma\tau} \nabla^{\sigma}   \hat{T}_\nu{}^{\tau}
-\frac1{\tau^2}\, \epsilon_{\mu\sigma\tau} \epsilon_{\nu\kappa\lambda}\,\mathbb{S}^\kappa{}^\sigma \hat{T}^\lambda{}^\tau
-\frac1{2\tau^2}\, \epsilon_{\mu\sigma\tau} \epsilon_{\nu\kappa\lambda}\,\hat{T}^\kappa{}^\sigma \hat{T}^\lambda{}^\tau
-\frac1{\kappa}\,U_{\mu\nu}
\;,
\label{defT}
\end{equation}
in terms of the tensors (\ref{TU}),
where we have defined
\begin{equation}
\hat{T}_{\mu\nu}\coloneqq T_{\mu\nu}-\frac12\,g_{\mu\nu}\,T_\rho{}^\rho
\;.
\end{equation}
Equations (\ref{eom_T}) are still third way consistent in the above sense, i.e.\ their divergence vanishes upon iterating the equation,
if the source tensor satisfies the identity~\cite{Arvanitakis:2014yja}
\begin{equation} 
\nabla^\mu \mathbb{T}_{\mu\nu} =
\frac1{\tau}\,\epsilon_{\nu\rho\sigma}\,\mathbb{S}^{\kappa\rho}\, \mathbb{T}_{\kappa}{}^{\sigma}
\;.
\label{divTT}
\end{equation}
Using the explicit expression (\ref{defT}) for the source tensor, after some computation and iterating the field equations, 
this condition reduces to 
\begin{equation}
\nabla^\mu U_{\mu\nu} =
\frac{1}{\tau}\,\epsilon_{\nu\rho\kappa}  
\left(\mathbb{S}^{\rho\mu}+\hat{T}^{\rho\mu}\right)U_\mu{}^{\kappa}
-\epsilon_{\nu\kappa\lambda}   {T}^{\kappa\lambda} 
\;,
\label{cond1U}
\end{equation}
in terms of the tensor $T_{\mu\nu}$ and $U_{\mu\nu}$. 
For bosonic matter, this condition is identically satisfied since $U_{\mu\nu}$ vanishes and $T_{\mu\nu}$ is symmetric.
A second consistency condition for (\ref{eom_T}) arises from requiring the source tensor $\mathbb{T}_{\mu\nu}$ to be symmetric
which is not manifest from its definition (\ref{defT}). Explicitly, this reduces to the condition 
\begin{equation}
\nabla^{\mu} {T}_{\mu\nu}
=
-\frac1{\tau}\, \epsilon_{\rho\kappa\lambda}\,\left(\mathbb{S}^\rho{}_\nu+\hat{T}^\rho{}_\nu \right){T}^{\kappa\lambda} 
-\frac{\tau}{\kappa}\,\epsilon_{\nu\kappa\lambda}\,U^{\kappa\lambda}
\;.
\label{cond0U}
\end{equation}
Again, for vanishing $U_{\mu\nu}$ this simply reduces to $T_{\mu\nu}$ being symmetric and conserved.
For non-vanishing $U_{\mu\nu}$, i.e.\ non-trivial fermion couplings to the spin connection $\varpi_\mu{}^a$
on the other hand, consistency of the field equations (\ref{eom_T}) requires the couple of consistency conditions (\ref{cond1U}), (\ref{cond0U}).
It is instructive to see explicitly how these consistency conditions indeed arise as a consequence of Lorentz and diffeomorphism symmetry
of the Lagrangian (\ref{matter coupling}).
To this end, we  note that Lorentz invariance of ${\cal L}_{\rm matter}$ gives rise to
\begin{equation}
0 = \delta_\Lambda {\cal L}_{\rm matter} =
2\,\sqrt{-g}\left(
\varepsilon^{abc} e_{\mu c} \Lambda_a T^\mu{}_b +D[\varpi]_\mu \Lambda^a U^\mu{}_a \right)
+ \delta_\Lambda \Phi \,\frac{\partial{\cal L}_{\rm matter}}{\partial \Phi}
\;,
\label{invLorentz}
\end{equation}
where we have used (\ref{Lorentz}). For the last term, we may use that
$\frac{\partial{\cal L}_{\rm matter}}{\partial \Phi}=\frac{\partial{\cal L}}{\partial \Phi}$,
showing that this term vanishes on-shell.
The remaining part of (\ref{invLorentz}) then reproduces (\ref{cond1U}) after using the equations of motion, and up to higher order fermion terms.
Similarly, one finds that diffeomorphism invariance of (\ref{matter coupling}) implies the condition (\ref{cond0U}).

Let us finally comment on the form (\ref{defT}) of the source tensor $\mathbb{T}_{\mu\nu}$. Surprisingly, at first glance, it does not appear to
carry a leading term linear in the standard energy-momentum tensor $T_{\mu\nu}$. On the other hand, such a contribution should naturally arise  
if equations (\ref{eom_final}) are considered as a deformation of a standard Lagrangian theory, for which the source tensor is the energy-momentum tensor,
c.f.\ the constructions of~\cite{Arvanitakis:2014yja,Ozkan:2018cxj,Cebeci:2020amh}. In fact, such a term can easily be generated in (\ref{defT}) upon including 
in ${\cal L}_{\rm matter}$ a further contribution to the cosmological constant according to
\begin{equation}
{\cal L}_{\rm matter}[e, \varpi, \Phi] \longrightarrow {\cal L}_{\rm matter}[e, \varpi, \Phi] + \mu_2\,\sqrt{-g}
\;.
\end{equation}
In presence of this term, the source tensor (\ref{defT}) changes according to
\begin{equation}
\mathbb{T}_{\mu\nu}
\longrightarrow
\mathbb{T}_{\mu\nu}
+\frac{\mu_2}{4\tau^2}\,T_{\mu\nu}
+\frac{\mu_2}{4\tau^2}\,\mathbb{G}_{\mu\nu}
+\frac{\mu_2^2}{16\tau^2}\,g_{\mu\nu}
\;,
\nonumber\\
\end{equation}
and now exhibits an explicit term in $T_{\mu\nu}$. The resulting field equations reproduce the constructions of~\cite{Arvanitakis:2014yja,Ozkan:2018cxj}. 
In section~\ref{subsec:TMSG} below we will explicitly see how this mechanism is at work in taking the limit by which the 
(matter coupled) MMG equations reduce to the (matter coupled) TMG equations.

%%%%%%%%%%%%%%%%%%%%%%%%%%%%%%%%%%%%%%%%%%%%%%%%%%%%%%%
%%%%%%%%%%%%%%%%%%%%%%%%%%%%%%%%%%%%%%%%%%%%%%%%%%%%%%%

\section{Supersymmetry}
\label{sec:susy}

%%%%%%%%%%%%%%%%%%%%%%%%%%%%%%%%%%%%%%%%%%%%%%%%%%%%%%%

We have in the previous section introduced a new class of bosonic actions that describe the dynamics of minimal massive gravity and
higher extensions thereof.
In this section, we focus on the MMG Lagrangian (\ref{LMMG}) and construct its minimal supersymmetric extension up to and
including quartic fermion terms. 
Different parts of this Lagrangian have been embedded into supersymmetric models in the context of 
super Chern-Simons theories \cite{Achucarro:1987vz}, topologically massive supergravity~\cite{Deser:1982sw,Deser:1984cts},
and first order formulations thereof~\cite{Giacomini:2006dr,Sezgin:2009dj,Routh:2013uc}.
The resulting theory here will carry two gravitino fields, $\{\psi_\mu, \Psi_\mu\}$, transforming however under the same local supersymmetry parameter $\epsilon$
as
\begin{equation}
\delta_\epsilon \psi_\mu = D[\mathring\omega]_\mu \epsilon + \dots \;,\qquad
 \delta_\epsilon \Psi_\mu = D[\varpi]_\mu \epsilon + \dots \;.
\end{equation}
This is the fermionic analogue of the fact that the bosonic sector of (\ref{LMMG}) carries the dreibein $e_\mu{}^a$ with associated Levi-Civita connection $\mathring\omega_\mu{}^a$ together with an independent spin connection $\varpi_\mu{}^a$.
As a result, the fermionic sector carries a massive spin 3/2 mode which constitutes the superpartner to the massive spin-2 mode.
We will confirm this in detail when analyzing the theory around given AdS vacua.

Before presenting the result, let us introduce our spinor conventions. We use $(-++)$ signature for the metric, and $D=3$ gamma matrices with
 \begin{equation}
\gamma_{\mu\nu} = \epsilon_{\mu\nu\rho}\,\gamma^\rho
\;.
\end{equation}
All spinors are Majorana, such that
 \begin{equation}
 \bar\epsilon \chi = \bar\chi \epsilon
 \;,\qquad
 \bar\epsilon\gamma_\mu \chi = -\bar\chi \gamma_\mu \epsilon
 \;.
\end{equation}
Covariant derivatives with respect to different spin connections are defined as
\begin{equation}
D[\omega]_\mu \epsilon =  \partial_\mu \epsilon+\tfrac12\omega_\mu{}^a \gamma_a\,\epsilon \;,\quad
D[\varpi]_\mu \epsilon  =  \partial_\mu \epsilon+\tfrac12\varpi_\mu{}^a \gamma_a\,\epsilon \;,\quad \mbox{etc.}
\;.
\end{equation}

\subsection{The supersymmetric Lagrangian}

Our ansatz for the supersymmetric extension of MMG is the following Lagrangian
\begin{align}
{\cal L}[e,\varpi,\psi,\Psi] =\;&
\varepsilon^{\mu\nu\rho}\Big(e_\mu{}^{a}R[\omega]_{\nu\rho, a}
+\lambda\, \varepsilon_{a b c} e_\mu{}^{a} e_\nu{}^{b} e_\rho{}^{c}
+ \tau\,e_\mu{}^{a}{} D[\varpi]_\nu  e_{\rho a} \Big)
+\kappa\,{\cal L}_{\rm CS}[\varpi] 
\nonumber\\
&{}\,
-\varepsilon^{\mu\nu\rho}
\bar\psi{}_\mu D[\omega]_\nu \psi{}_\rho
  +\frac{1}4\left(\eta\,\tau+\tfrac1{\eta\,\kappa}\right)
  \varepsilon^{\mu\nu\rho}
 \,\bar\psi{}_\mu \gamma_{\nu} \psi_\rho 
\nonumber\\
&{}\,
+\frac1{\eta}\,\varepsilon^{\mu\nu\rho}
\, \bar\Psi{}_\mu D[\varpi]_\nu \Psi{}_\rho
  -\frac{\tau}2\,
  \varepsilon^{\mu\nu\rho}
 \,\bar\Psi{}_\mu \gamma_{\nu} \Psi_\rho 
  +\tau\,\varepsilon^{\mu\nu\rho}
 \,\bar\chi{}_\mu \gamma_{\nu} \chi_\rho 
\;.
\label{LMMSG}
\end{align}
In the last term, we have introduced the combination
\begin{equation}
\chi_\mu = \Psi_\mu -\psi_\mu
\;,
\label{chi}
\end{equation}
capturing the difference between the two gravitino fields.
The new parameter $\eta$ featuring in the fermionic couplings of (\ref{LMMSG}) is 
related to the bosonic coupling constants $\{\kappa, \tau, \lambda\}$ by the relation\footnote{
In \cite{Deger:2022gim} we used notation in which $\eta$ was represented as $\eta=\zeta^2$ which however hides the fact that $\eta$ can take negative values.}
\begin{equation}
\lambda
=
\frac1{12}\left(\eta\,\tau+\tfrac1{\eta\,\kappa}\right)^2
-\frac{\tau}{3\,}\left(\eta \, \tau-\tfrac{1}{\eta\,\kappa}\right)
.
\label{lambda_eta0}
\end{equation}
It is invariant under the scaling symmetry (\ref{scaling}).
An MMG model with given bosonic coupling constants  $\{\kappa, \tau, \lambda\}$ thus allows the supersymmetrization (\ref{LMMSG}) if
equation (\ref{lambda_eta0}) admits real solutions for $\eta$. 
We will discuss the full landscape of supersymmetric theories and their vacua in section~\ref{sec:AdSvacua} below.
We shall refer to the model (\ref{LMMSG}) as minimal massive supergravity (MMSG).

The Lagrangian (\ref{LMMSG}) is considered as a second order Lagrangian in the dreibein $e_\mu{}^a$, with the spin connection
$\omega_\mu{}^a$ determined as a function of $e_\mu{}^a$, by means of the torsion equation
\begin{equation}
D[\omega]_{[\mu} e_{\nu]}{}^a = -\frac{1}{4}\,\psi_\mu \gamma^{a} \psi_\nu
\;,
\end{equation}
as customary in the `1.5 order formalism'. Consequently, the spin connection $\omega_\mu{}^a$ has 
non-vanishing torsion $T[\omega]_\mu{}^a$ and contorsion $K[\omega]_\mu{}^a$, bilinear in the fermion fields,
and can be given explicitly as
\begin{align}
\omega_\mu{}^a =&\, \mathring{\omega}_\mu{}^a + K[\omega]_\mu{}^a
\nonumber\\
=&\,
\mathring{\omega}_\mu{}^a 
-\tfrac{1}4\,\epsilon^{\rho\sigma\tau}\, e_\rho{}^{a}  \, 
\bar\psi_\sigma \gamma_{\mu} \psi_\tau
+\tfrac{1}8\, \epsilon^{\rho\sigma\tau} \,e_\mu{}^{a}  \,
\bar\psi_\sigma \gamma_{\rho} \psi_\tau
\;,
\label{omega}
\end{align}
in terms of the torsionless Levi-Civita connection $\mathring{\omega}_\mu{}^a$.

The fermionic field equations are obtained by variation 
\begin{equation}
\delta {\cal L} =
-2\,\delta \bar\psi_\mu \, {\cal E}^\mu
+\frac2{\eta}\,\delta \bar\Psi{}_\mu\,  {\cal E}_+^\mu
 \;,
\end{equation}
and read
\bea
{\cal E}^\rho &=&
\varepsilon^{\mu\nu\rho}
 \, \left( D[\omega]_\mu\psi_\nu
 -\frac14\left(\eta\,\tau+\tfrac{1}{\eta\,\kappa}\right) \gamma_{\mu} \psi_\nu
 +\tau\,\gamma_{\mu} \chi_\nu
 \right)
 ,
\nonumber\\
{\cal E}_+^\rho&=&
\varepsilon^{\mu\nu\rho}
 \, \left( D[\varpi]_\mu \Psi_\nu
-\frac12\,\eta\,\tau\, \gamma_{\mu} (\Psi_\nu-2\chi_\nu)
    \right)
    .
    \label{eom_ferm_exp}
\eea

For the supersymmetry transformations of the various fields of (\ref{LMMSG}) we start from the following ansatz
\begin{align}
\delta_\epsilon e_\mu{}^{a} =\;&
\frac{1}{2}\,\bar\psi_\mu \gamma^{a} \epsilon
\;,\nonumber\\
\delta_\epsilon \psi_\mu =\;& D[\omega]_\mu \epsilon 
-\frac14 \left(\eta\tau+\tfrac1{\eta\,\kappa}\right) \gamma_\mu \epsilon
\;,\nonumber\\
\delta_\epsilon \varpi_\mu{}^{a} =\;&
-\frac1{2\eta\kappa}\,\bar\Psi_\mu  \gamma^{a} \epsilon
-\frac{1}{2}\,D[\varpi]_\mu \left(\bar\chi_\nu \epsilon \,e^{\nu a}\right)
,\nonumber\\
\delta_\epsilon \Psi_\mu =\;&
D[\varpi]_\mu \epsilon -\frac12\,\eta\tau\, \gamma_\mu \epsilon
+\frac{1}{4}\,(\bar\chi_\lambda \epsilon) \,\gamma^{\lambda}\,\Psi_\mu
\;,
\label{susy_full}
\end{align}
which is complete to quadratic order in the fermions but will receive further higher order fermion contributions,
although some cubic fermion terms are already contained in $\delta_\epsilon \psi_\mu$ (via (\ref{omega})) and $\delta_\epsilon \Psi_\mu$.

In order to prove invariance of the Lagrangian (\ref{LMMSG}) under supersymmetry (\ref{susy_full}), it is helpful to properly organize the Lagrangian into
its different parts. Note that upon rescaling
\bea
\eta\tau=\tilde{\tau}\;,\quad
\eta\kappa=\tilde{\kappa}
\;,
\eea
the supersymmetry transformations (\ref{susy_full}) no longer depend on $\eta$, while the Lagrangian (\ref{LMMSG}) falls into
two parts of order $\eta^0$ and $\eta^{-1}$ which ought to be separately supersymmetric.\footnote{We thank Mehmet Ozkan for pointing this out. A similar
structure is exhibited by the second order Lagrangian of TMSG \cite{Gibbons:2008vi}.}
This observation motivates rewriting of the Lagrangian (\ref{LMMSG}) as a sum of three terms
\begin{equation}
{\cal L}[e,\varpi,\psi,\Psi] =
{\cal L}_1 + {\cal L}_2 + {\cal L}_3
\;,
\label{LLL123}
\end{equation}
with
\begin{align}
{\cal L}_1[e,\psi] =\,&
\varepsilon^{\mu\nu\rho}\left(
e_\mu{}^{a}R[\omega]_{\nu\rho, a}
-
\bar\psi{}_\mu D[\omega]_\nu \psi{}_\rho
  -\frac12\,\mm\,\bar\psi{}_\mu \gamma_{\nu} \psi_\rho 
+\frac13\,\mm^2\,\varepsilon_{a b c} e_\mu{}^{a} e_\nu{}^{b} e_\rho{}^{c}
\right)
,
\nonumber\\[1ex]
{\cal L}_2[e,\varpi,\Psi] =\,&
 \tau\,\varepsilon^{\mu\nu\rho}\left(
    e_\mu{}^{a}{} D[\varpi]_\nu  e_{\rho a} 
     -\frac{2}{3}\,\beta\,
\varepsilon_{a b c} e_\mu{}^{a} e_\nu{}^{b} e_\rho{}^{c} \right)
+\kappa\,{\cal L}_{\rm CS}[\varpi]      
+\frac1{\eta}\,\varepsilon^{\mu\nu\rho}
\, \bar\Psi{}_\mu D[\varpi]_\nu \Psi{}_\rho\nonumber\\
&\quad
  -\frac{\tau}2\,
  \varepsilon^{\mu\nu\rho}
 \,\bar\Psi{}_\mu \gamma_{\nu} \Psi_\rho 
 \;,
\nonumber\\[2ex]
{\cal L}_3[e,\psi,\Psi] =\,&\tau\,\varepsilon^{\mu\nu\rho}
 \,\bar\chi{}_\mu \gamma_{\nu} \chi_\rho 
 \;,
 \label{L123}
 \end{align}
where we have introduced the combinations of parameters
\begin{equation}
\mm = -\frac{1}2\left(\eta\,\tau+\frac1{\eta\,\kappa}\right)
,\quad\quad
\beta=\frac{1}2\left(\eta\,\tau-\frac1{\eta\,\kappa}\right)
.
\label{mbeta}
\end{equation}
Note that, we have  $3\lambda = m^2-2\tau \beta$ from \eqref{lambda_eta0} and $m^2 -\beta^2 = \tau/\kappa$.
The first part ${\cal L}_1$ of (\ref{L123}), which only depends on $e_\mu{}^a$ and $\psi_\mu$, is precisely the 
${\cal N}=(1,0)$ supersymmetric extension of 2+1 AdS gravity \cite{Achucarro:1987vz}. Indeed, it is by itself invariant under
the relevant part of the supersymmetry variation (\ref{susy_full})
\begin{equation}
\delta_\epsilon e_\mu{}^{a} =
\frac{1}{2}\,\bar\psi_\mu \gamma^{a} \epsilon
\;,\quad\quad
\delta_\epsilon \psi_\mu = D[\omega]_\mu \epsilon 
+\frac{\mm}2\, \gamma_\mu \epsilon
\;,
\end{equation}
to all order in fermions.
The second part ${\cal L}_2$ of (\ref{L123}) corresponds to the supersymmetric extension \cite{Giacomini:2006dr}
of the exotic action of 2+1 AdS gravity \cite{Witten:1988hc}. 
In order to study its behavior under the supersymmetry transformations (\ref{susy_full}), let us rewrite these
transformations into the form
\begin{align}
\delta_\epsilon e_\mu{}^{a} =\;&
\frac{1}{2}\,\bar\Psi_\mu \gamma^{a} \epsilon
\textcolor{red}{{}-{}\frac{1}{2}\,\bar\chi_\mu \gamma^{a} \epsilon
{}+{}\frac{1}{2}\,\bar\chi_\lambda \epsilon \,
\varepsilon^{abc}\,e_{\mu b}\,e_{c}{}^{\lambda}}
\textcolor{blue}{{}-{}\frac{1}{2}\,\bar\chi_\lambda \epsilon \,
\varepsilon^{abc}\,e_{\mu b}\,e_{c}{}^{\lambda}}
\;,\nonumber\\
\delta_\epsilon \varpi_\mu{}^{a} =\;&
-\frac1{2\eta\kappa}\,\bar\Psi_\mu  \gamma^{a} \epsilon
\textcolor{blue}{{}-{}\frac{1}{2}\,D[\varpi]_\mu \left(\bar\chi_\nu \epsilon \,e^{\nu a}\right)}
\;,\nonumber\\
\delta_\epsilon \Psi_\mu =\;&
\left(D[\varpi]_\mu -\frac12\,\eta\tau\, \gamma_\mu\right) \epsilon
\textcolor{blue}{{}+{}\frac{1}{4}\,(\bar\chi_\lambda \epsilon) \,\gamma^{\lambda}\,\Psi_\mu}
\;.
\label{susy_2p}
\end{align}
It is straightforward to check that the first terms (indicated with black) in (\ref{susy_2p}) define an invariance of the Lagrangian ${\cal L}_2$ to all orders in fermions, 
corresponding to a special case ($\alpha_1=0$) of \cite{Giacomini:2006dr}.
In turn, the last terms (indicated with \textcolor{blue}{blue}) in (\ref{susy_2p}) constitute a Lorentz transformation and thus a separate invariance of the Lagrangian ${\cal L}_2$.
Supersymmetry variation of ${\cal L}_2$ thus reduces to the contributions from the \textcolor{red}{red} terms in (\ref{susy_2p}) (two middle terms in the first equation)
which yield
\begin{align} 
\delta_\epsilon {\cal L}_2=&\,
\tau\,
\varepsilon^{\mu\nu\rho}\,(\bar\chi_\rho \epsilon) \,K[\varpi]_{\mu\nu}
-\tau\,e\,(\bar\chi_\mu \gamma^{\nu} \epsilon)\,K[\varpi]_\nu{}^\mu
+\tau\,e\, (\bar\chi_\mu \gamma^{\mu} \epsilon) \left( K[\varpi]_\nu{}^\nu -2\,\beta \right)
      \nonumber\\
&  
+\frac{\tau}4\,
  \varepsilon^{\mu\nu\rho}
 \,(\bar\Psi{}_\mu \gamma_{a} \Psi_\rho)(\bar\chi_\nu \gamma^{a} \epsilon)
-\frac{\tau}2\,e
 \,(\bar\Psi{}_\rho \gamma^{\mu} \Psi_\mu)(\bar\chi^\rho \epsilon)
 \;.
 \label{L2var}
\end{align}
These terms must be compensated by the variation of the remaining term ${\cal L}_3$ in (\ref{LLL123})
which can be spelled out as
\begin{align} 
\delta_\epsilon {\cal L}_3 =&\,  
-\tau\,
\varepsilon^{\mu\nu\rho}\,(\bar\chi_\rho \epsilon) \,K[\varpi,\omega]_{\mu\nu}
+\tau\,e\,(\bar\chi_\mu \gamma^{\nu} \epsilon)\,K[\varpi,\omega]_\nu{}^\mu
%      \nonumber\\
%&  
-\tau\, e\,(\bar\chi_\mu \gamma^{\mu} \epsilon) \left( K[\varpi,\omega]_\nu{}^\nu -2\,\beta \right)
      \nonumber\\
&  
+\frac{\tau}{2}\,\varepsilon^{\mu\nu\rho} \, (\bar\chi_\nu \epsilon) \,(\bar\chi_\mu \Psi_\rho)
+\frac{\tau}{2}\,\varepsilon^{\mu\nu\rho} (\bar\chi_\mu \gamma_a \chi_\rho)\,(\bar\psi_\nu \gamma^{a} \epsilon)
+\frac{\tau}{2}\,e\,(\bar\chi_\mu \epsilon) \, (\bar\chi^\mu \gamma^{\rho}\,\Psi_\rho)
      \nonumber\\
&  
+\frac{\tau}{2}\,e\,(\bar\chi_\mu \epsilon) \,  (\bar\Psi^\mu \gamma^{\rho}\,\chi_\rho)
\;,
 \label{L3var}
\end{align}
with
\begin{equation}
K[\varpi,\omega]_{\mu}{}^a = K[\varpi]_{\mu}{}^a-K[\omega]_{\mu}{}^a
\;.
\label{Kww}
\end{equation}
Upon using the explicit expression (\ref{omega}) for the contorsion $K[\omega]_{\mu}{}^a$,
the sum of (\ref{L2var}) and (\ref{L3var}) reduces to terms quartic in the fermionic fields.
With the Fierz identities 
\begin{align} 
\varepsilon^{\mu\nu\rho}
 \,(\bar\chi_\mu \gamma_\sigma \chi_\nu )\, (\bar\chi_\rho\gamma^\sigma \epsilon )
=&\, 
0\;,
\nonumber\\
 (\bar\chi_{\rho}\gamma^\rho \chi_\sigma)\,(\bar\epsilon\gamma^\sigma \gamma^{\nu} \chi_\nu)
=&\, 
0
\;,
 \nonumber\\
\varepsilon^{\rho\sigma\tau}\,(\bar\chi_{\rho}\gamma_\sigma\chi_\tau)\,
 (\bar\epsilon\gamma^\nu \chi_\nu) 
=&\, 
  (\bar\chi^\rho \chi_\rho)\,(\bar\epsilon \gamma^\nu \chi_\nu)
  =
  -2\,(\bar\chi^\mu \epsilon )
\, (\bar\chi{}_\mu  \gamma^\nu \chi{}_\nu)
\;,
\label{Fierz}
\end{align}
we arrive at the final result
\begin{equation} 
\delta_\epsilon {\cal L} = 
\delta_\epsilon {\cal L}_2+\delta_\epsilon {\cal L}_3
=
\frac{\tau}{2}\,(\bar\chi_\mu \epsilon) \, (\bar\chi^\mu \gamma^{\nu}\chi_\nu)
\;.
\label{final-4F}
\end{equation}
While it is remarkable that the only remaining contribution of the variation
is a single term cubic in the combination $\chi_\mu$ of (\ref{chi}), the presence of this term a priori still poses
an obstruction to supersymmetry at higher fermion order.
We observe however, that such a term can precisely be cancelled by the contributions coming from an additional 
quartic fermion term in the Lagrangian
\begin{equation}
{\cal L}_{\chi^4} = e\,(\epsilon^{\mu\nu\rho} \bar\chi_\mu \gamma_\nu \chi_\rho)^2
\;,
\label{Lchi4}
\end{equation}
whose variation explicitly reads
\bea
\delta {\cal L}_{\chi^4} &=&  
-\frac1{\tau}\,(\varepsilon^{\mu\nu\lambda} {\cal E}_{\mu\nu, a})
\,e_{\lambda b}\,\varepsilon^{abc}\,e_{c}{}^{\rho}\,\bar\epsilon_3 \Psi_\rho
-\frac4{\eta\tau}\,\bar\epsilon_3 D[\varpi]_\rho {\cal E}^\rho_+ 
+ \frac1{\eta\tau}\,\varepsilon^{\mu\nu\rho}
 \,  \bar{\cal E}_{\mu\nu}{}^{a} \,\bar\epsilon_3 \gamma_{a}  \Psi_\rho
\nonumber\\
&&{}
-
2 \,\bar\epsilon_3 \gamma_{\rho}   {\cal E}_+^{\rho}
+4\,\tau\,\bar\epsilon_3 \gamma_\rho {\cal E}^\rho 
-8\,(\beta-\tau)\, (\bar\epsilon \gamma^\sigma \chi_\sigma)(\varepsilon^{\mu\nu\rho} \bar\chi_\mu \gamma_\nu \chi_\rho)
+\dots
\;,
\label{varChi4}
\eea
up to terms of order six in the fermions, 
with field equations ${\cal E}_{\mu\nu, a}, \bar{\cal E}_{\mu\nu, a}$ from 
(\ref{eom_bosonic}), and ${\cal E}^\rho, {\cal E}^\rho_+$ from (\ref{eom_ferm_exp}).
The parameter $\epsilon_3$ in (\ref{varChi4}) is given by
\begin{equation}
\epsilon_3= (\epsilon^{\mu\nu\rho} \bar\chi_\mu \gamma_\nu \chi_\rho)\,\epsilon\;.
\end{equation}
In deriving this variation, we have used the relation
\bea
\frac2{\eta\tau}\,D[\varpi]_\rho {\cal E}^\rho_+ &=&
\frac1{2\eta\tau}\,\varepsilon^{\mu\nu\rho}
 \,  \bar{\cal E}_{\mu\nu}{}^{a} \,\gamma_{a}  \,\Psi_\rho
-
 \,\gamma_{\rho}   {\cal E}_+^{\rho}
+2\,\tau\,\gamma_\rho {\cal E}^\rho 
%  \nonumber\\
%&&{}
-\frac1{2\tau}\,(\varepsilon^{\mu\nu\lambda} {\cal E}_{\mu\nu, a})
\,e_{\lambda b}\,\varepsilon^{abc}\,e_{c}{}^{\rho}\,\psi_\rho
\nonumber\\
&&{}
+
 \,  K[\varpi,\omega]_\mu{}^{\nu}\,\gamma^\mu  \chi_\nu
- K[\varpi,\omega]_{\nu}{}^{\nu} \,  \gamma^\rho \chi_\rho
+2\,(2\,\tau-\beta)\, \gamma^\nu \chi_\nu
+ \dots
\;,
 \label{DEF1}
\eea
among the fermionic and the bosonic field equations (\ref{eom_bosonic}), (\ref{eom_ferm_exp}),
up to cubic fermion terms. 
As a result, the variation (\ref{varChi4}) thus reproduces a term proportional to (\ref{final-4F}) up to terms
proportional to the field equations, however dressed with functions that are at least cubic in the fermions.
All these terms can therefore be removed by adding proper higher order fermion terms to the 
supersymmetry transformations (\ref{susy_full}).
In presence of (\ref{Lchi4}), the Lagrangian is then invariant under supersymmetry at least 
up to and including quartic fermion terms. 
Potential terms of higher order in the fermions might require 
additional quintic corrections to (\ref{susy_full}). An all-order result would probably require the identification 
of some additional underlying structure, such as an extended supersymmetry or an off-shell formulation with further auxiliary fields 
in order to organize the higher order fermion contributions.

\subsection{Minimal massive supergravity (MMSG)}

We have shown in section~\ref{sec:bosonic} above, that the bosonic MMG equations (\ref{MMG0}) are obtained from the second order Lagrangian (\ref{LMMG}), after elimination of the connection $\varpi$ by its field equations. 
The resulting field equations are expressed exclusively in terms of the dreibein $e_\mu{}^a$, its Ricci tensor, and derivatives thereof.
We can now carry out the same computation starting from the full supersymmetric Lagrangian (\ref{LMMSG})
in order to obtain field equations in terms of the dreibein $e_\mu{}^a$ and its superpartner $\psi_\mu$.

The full bosonic field equations obtained from variation of (\ref{LMMSG}) read
\begin{align}
0 =&
R[\omega]_{\mu\nu,a}
+ \tau\,T[\varpi,\omega]_{\mu\nu,a}
+3\,\lambda\,\varepsilon_{abc}\,e_\mu{}^{b} e_\nu{}^{c}
  +\frac{\mm-\tau}2\, \bar\psi{}_\mu \gamma_{a}  \psi_\nu 
  +\frac{\tau}2 \,\bar\Psi{}_{\mu}  \gamma_{a}  \Psi_{\nu} 
  -\tau \,\bar\chi{}_{\mu} \gamma_{a}  \chi_{\nu}
 \;,\nonumber\\[1ex]
0 =& 
R[\varpi]_{\mu\nu,a} 
+ \frac{\tau}{\kappa}\,\varepsilon_{abc}\,
    e_\mu{}^{b}{}   e_{\nu}{}^{c}   
     - \frac1{2\,\eta\,\kappa}\,\bar\Psi{}_\mu \gamma_{a} \Psi_\nu 
     \;,
    \label{eom_bosonic_full}
\end{align}
up to quartic fermion terms and with torsionful spin connection $\omega_\mu{}^a$, c.f.\ (\ref{omega}). 
In analogy to (\ref{Kww}), we have introduced
\begin{equation}
T[\varpi,\omega]_{\mu\nu,a} = T[\varpi]_{\mu\nu,a}-T[\omega]_{\mu\nu,a}
\;.
\end{equation}
Solving the first equation of (\ref{eom_bosonic_full}) for $K[\varpi,\omega]$ and plugging all into
(\ref{RRKK}), straightforwardly yields the fermionic extension of the MMG equations (\ref{MMG0}) to quadratic order in the fermions.

This extension is a particular example of the general matter coupling discussed in section~\ref{subsec:matter} above.
The MMG equations are modified according to (\ref{eom_T}) by a source tensor (\ref{defT})
based on the particular energy-momentum tensor
\bea
T^{\mu\nu} &=&
\frac12\,\epsilon^{\lambda\sigma(\mu}\epsilon^{\nu)\rho\tau}   \,  \bar\psi_\rho \gamma_\lambda \nabla[\omega]_{\sigma}\psi_\tau
+ \frac{m}4\,\varepsilon^{\sigma\tau(\mu} \,\bar\psi{}_\sigma \gamma^{\nu)} \psi_\tau 
  +\frac{\tau}4\,\varepsilon^{\sigma\tau(\mu} \,\bar\Psi{}_\sigma \gamma^{\nu)} \Psi_\tau 
  -\frac{\tau}{2}\,\varepsilon^{\sigma\tau(\mu}\,\bar\chi{}_\sigma \gamma^{\nu)} \chi_\tau
\nonumber\\
&&{}
-\frac{\tau}4\,\epsilon^{\mu\nu\lambda} \,  \bar\psi_\lambda \gamma^\kappa \psi_\kappa
+\frac{\tau}4\,\epsilon^{\mu\nu\lambda}  \,  \bar\chi_\lambda \gamma^\kappa \chi_\kappa
-\frac{\tau}{2}\,\bar\psi^{[\mu} \chi^{\nu]}
\;,
\label{TMMSG}
\eea
bilinear in the fermions. Moreover, the tensor $U_{\mu\nu}$ defined in \eqref{TU} entering the source tensor is given by
\begin{equation}
U_{\mu\nu} = 
\frac1{4\eta}\,\epsilon_{\mu\sigma\tau} \,\bar\Psi^\sigma \gamma_\nu \Psi^\tau
\;.
\label{UMMSG}
\end{equation}
As a final step, we may eliminate the auxiliary fermion field $\chi_\mu$ from these expressions
upon solving the first of the field equations ${\cal E}^\rho$ from (\ref{eom_ferm_exp}) as
\bea
\chi_\mu &=& 
 -\frac{\mm}{2\tau}\,\psi_\mu - \frac{1}{2\tau}\,\gamma_\nu\gamma_\mu R^\nu
\;,
\label{chipsi}
\eea
for $\chi_\mu$ in terms of the gravitino $\psi_\mu$ and the gravitino curvature
\begin{equation}
R^\mu = \epsilon^{\mu\nu\rho}\,D[\omega]_\nu \psi_\rho
\;.
\end{equation}
Plugging this into (\ref{TMMSG}), (\ref{UMMSG}) and everything back into (\ref{defT}), finally yields the fermionic completion (\ref{eom_T})  of the MMG equations, expressed only in terms of the gravitino $\psi_\mu$ and its derivatives.
The full equations continue to be third-way consistent as discussed in section~\ref{subsec:matter}.

In a similar fashion, we can obtain the fermionic equations of motion as higher order equations exclusively in terms 
of the dreibein $e_\mu{}^a$, the gravitino $\psi_\mu$, and their derivatives. To this end, we plug (\ref{chipsi}) into
the second field equation ${\cal E}_+^\rho$  of (\ref{eom_ferm_exp}), and obtain after some computation 
\begin{align}
\tau\,C^\rho=&\,
\frac1{2}\,\Big((\mm-2\tau)^2-4\eta\,\tau^2\Big)
\,R^{\rho}
+\frac{\mm}{2}\, \epsilon^{\mu\nu\rho}
\,\gamma_\sigma \psi_\nu\, S_\mu{}^\sigma \,
+\frac{1}{4}
\mm\,\Big((\mm-2\tau)^2-4\eta\,\tau^2\Big)\,
 \epsilon^{\mu\nu\rho}\, \gamma_{\mu} \psi_\nu
\nonumber\\
&{}
-\frac1{2}
\, R^{\mu}\, G_\mu{}^\rho \,
-\frac1{2}\,\epsilon^{\rho\mu\nu}
\,\gamma^\sigma R_\mu\, G_{\nu\sigma} \,
-\frac1{2}\,\epsilon^{\rho\mu\nu}
\gamma_\mu   R^{\sigma}\, G_{\nu\sigma} \,
\;,
\label{eom_fermion_psi}
\end{align}
up to cubic fermion terms.
Here, we have used (\ref{eom_bosonic_full}) to eliminate $K[\varpi]_\mu{}^{a}$,
and introduced the $\gamma$-traceless ``Cottino vector-spinor'' \cite{Gibbons:2008vi}
\begin{equation}
C^\mu = 
\gamma^\rho\gamma^{\mu\nu}\,D[\omega]_\nu R_\rho - \epsilon^{\mu\nu\rho}\,S_{\rho\sigma}\,\gamma^\sigma \psi_\nu
\;.
\label{Cmu}
\end{equation}
Equation (\ref{eom_fermion_psi}) constitutes the second-order fermionic field equation, 
which yields the ``super-partner'' to the bosonic MMG equations with fermionic sources as found above.

\subsection{TM(S)G limit}
\label{subsec:TMSG}

In this section, we study the limit in which the minimal massive supergravity constructed above
reduces to the supersymmetric extension of topologically massive gravity (TMSG) \cite{Deser:1982sw,Deser:1984cts}.
For the bosonic MMG equation in the original conventions of (\ref{MMG0}), the limit to the TMG equations \cite{Deser:1981wh}
corresponds to 
\begin{equation}
\gamma\rightarrow0\;,\quad
\bar\sigma\rightarrow\sigma\;,\quad
\bar\Lambda_0 \rightarrow \Lambda_0
\;,
\end{equation}
upon which the field equations reduce to
\begin{equation}
\frac1{\mu}\,C_{\mu\nu} +
\sigma\, G_{\mu\nu}
+ \Lambda_0 \, g_{\mu\nu}
=
0
\;.
\label{TMG}
\end{equation}
In terms of the parameters $\{\tau, \kappa, \lambda\}$ of our bosonic Lagrangian (\ref{LMMG}), 
this limit can conveniently be implemented by setting
\begin{equation}
\tau = - \frac{\mu \,(1+\alpha \sigma)^2}{\alpha} \, , \quad
\kappa = \frac{\alpha}{\mu}\, , \quad
\lambda = \frac13\,\alpha\,\Lambda_0-\frac{2\mu^2\,(1+\alpha\sigma)^3}{3\alpha^2}\, ,
\label{rule12}
\end{equation}
c.f.\ \eqref{identication}, and sending $\alpha \rightarrow 0$. 
Note that the TMG limit imposes $\kappa\tau<0$, as also follows directly from (\ref{param}) upon sending $\gamma\rightarrow0$.

In order to study this limit on the level of the Lagrangian, it is convenient to express the connection $\varpi_\mu{}^a$ as
\begin{equation}
\varpi_\mu{}^a = \omega_\mu{}^a -\frac{3\lambda}{2\,\tau}\,e_\mu{}^a + \alpha\,\beta_\mu{}^a
\;,
\label{varpiOmega}
\end{equation}
in terms of a new field $\beta_\mu{}^a$. Recall that although we have introduced the bosonic Lagrangian (\ref{LMMG}) as a second order Lagrangian ${\cal L}[e_\mu{}^a, \varpi_\mu{}^a]$, it may alternatively be considered as a first order Lagrangian with (\ref{L0MMG_Pal}) carrying an a priori independent spin connection $\omega_\mu{}^a$ which is then determined by its field equation. We can take the TMG limit for both formulations simultaneously. To this end, we plug the expansions (\ref{rule12}), (\ref{varpiOmega}) into the bosonic Lagrangian (\ref{LMMG}) and expand in $\alpha$. After some computation this leads to
\bea
{\cal L}_{\rm bos}&=&
\alpha\,\varepsilon^{\mu\nu\rho}\, \left(
-\sigma\,e_\mu{}^a R[\omega]_{\nu\rho\,a}
+\frac{\Lambda_0}{3}\,\varepsilon_{abc}\,
    e_\mu{}^{a}{} e_\nu{}^b  e_{\rho}{}^c \right)
+\frac1{\mu}\,{\cal L}_{\rm CS}[\omega]
\nonumber\\
&&{}
-2\,\alpha\,\varepsilon^{\mu\nu\rho}
\,\beta_\mu{}^{a} D[\omega]_\nu e_{\rho\,a}
~+{\cal O}(\alpha^2)
\nonumber\\[2ex]
&=&
\alpha\,{\cal L}_{\rm TMG}~+{\cal O}(\alpha^2)
\;.
\label{LTMGlimit}
\eea
Up to an overall scaling factor $\alpha$, this precisely reproduces the first order Lagrangian 
${\cal L}_{\rm TMG}$ of TMG \cite{Deser:1991qk,Carlip:1991zm} if the spin connection $\omega_\mu{}^a$ is considered as an independent field such that vanishing torsion is imposed by the Lagrange multiplier $\beta_\mu{}^a$. If instead $\omega_\mu{}^a$ is taken to be the torsionless Levi-Civita connection, the last term in (\ref{LTMGlimit}) identically vanishes, and the resulting Lagrangian is the second order TMG Lagrangian \cite{Deser:1981wh}.

In order to extend the computation to the full supersymmetric Lagrangian (\ref{LMMSG}), we find the expansion for the coupling constant $\eta$ from combining (\ref{lambda_eta0}) and (\ref{rule12}). Explicitly, we choose the branch
\begin{equation} \label{etatmg}
\eta = 1+\alpha\left(
\frac{m_0}{\mu} -\sigma \right) + {\cal O}(\alpha^2)
\;,
\qquad
m_0\coloneqq
\sqrt{-\Lambda_0/\sigma}\;,
\end{equation}
noting that the other branch does not allow for a smooth limit $\alpha\rightarrow0$.
This limit requires $\Lambda_0/\sigma<0$ which indeed is a necessary condition for
the supersymmetrizability of the TMG equations (\ref{TMG}).

In analogy to (\ref{varpiOmega}), we redefine the gravitino $\chi_\mu$ as
\begin{equation}
\chi_\mu =
 \frac{\alpha}{2\mu}
 \left(m_0\,\psi_\mu
+\gamma_\nu\gamma_\mu R^\nu\right)
+\alpha\,\theta_\mu
\;,
\label{chitheta}
\end{equation}
in terms of a new field $\theta_\mu$. Plugging this into the fermionic part of the Lagrangian (\ref{LMMSG}) we obtain after some computation
\bea
{\cal L}_{\rm ferm}
&=&
\alpha\,\Big( \sigma \,
 \bar\psi{}_\mu R^\mu
+\frac{1}{2\mu}
 \, \bar{R}^\nu \gamma_\mu \gamma_\nu R^\mu
  +\frac{\sigma m_0}{2}  \,
  \varepsilon^{\mu\nu\rho}
 \,\bar\psi{}_\mu \gamma_{\nu} \psi_\rho 
\nonumber\\
&&{}
\qquad
-\mu \,
\varepsilon^{\mu\nu\rho}
 \,{\bar\theta}{}_\mu \gamma_{\nu} \theta_\rho 
+\frac{1}{2}\, \varepsilon^{\mu\nu\rho}\,\beta_\nu{}^{a} 
 \bar\psi_\mu \gamma_{a}\psi_\rho \Big)
 +{\cal O}(\alpha^2)\;,
 \label{LTMSG}
\eea
where we have used among others the identity
\bea
\epsilon^{\mu\nu\rho}  \gamma_{\nu}  \gamma_\sigma\gamma_\rho &=&
2 \gamma_\sigma \gamma^\mu + 2\,\epsilon_\sigma{}^{\rho\mu}  \gamma_\rho
~=~2\,\delta_\sigma{}^\mu 
\;.
\eea
Indeed, this Lagrangian is precisely the supersymmetric extension of TMG in its first order formulation \cite{Sezgin:2009dj,Routh:2013uc}.
We may also perform the limit directly on the fermionic field equations (\ref{eom_fermion_psi}) which leads to the known super TMG equations
\bea
0
&=&
\frac1{2\sigma\mu}\,C^\rho
+R^{\rho} 
+\frac12\,m_0\, \gamma^{\rho\nu}
\,\psi_\nu\, 
\;,
\label{super-TMG}
\eea
with the Cottino vector-spinor from (\ref{Cmu}).
Note that in the TMG limit the supersymmetry transformation rules (\ref{susy_full}) reduce to
\begin{equation} \label{susytmg}
\delta_\epsilon e_\mu{}^{a} =
\frac{1}{2}\,\bar\psi_\mu \gamma^{a} \epsilon
\;,\qquad
\delta_\epsilon \psi_\mu = D[\omega]_\mu \epsilon 
+\frac{m_0}{2}\, \gamma_\mu \epsilon
\;,
\end{equation}
for $e_\mu{}^a$ and $\psi_\mu$. On the other hand, we may find the supersymmetry transformation of the fermionic auxiliary field $\theta_\mu$ by expanding the variation of (\ref{chitheta}) in $\alpha$. To lowest order in the fermions this leads to the transformation
\begin{equation}
\delta_\epsilon \theta_\mu =
-\frac1{4\mu}\,{\cal E}^{\rm (TMG)}_{\kappa\lambda,\mu}\,\epsilon^{\kappa\lambda\nu} \gamma_{\nu}    \epsilon 
+\frac1{8\mu}\,{\cal E}^{\rm (TMG)}_{\kappa\lambda,\nu}{}\,\epsilon^{\kappa\lambda\nu} \gamma_{\mu}    \epsilon 
\;,
\end{equation}
of $\theta_\mu$ into the bosonic field equations 
${\cal E}^{\rm (TMG)}_{\kappa\lambda,\mu}$ obtained as
\begin{equation}
\delta {\cal L}_{\rm TMG} =
\varepsilon^{\mu\nu\rho}\,\delta e_\mu{}^{a}\,{\cal E}^{\rm (TMG)}_{\nu\rho,a}
\;.
\end{equation}
This is consistent with the fermionic field equations $\theta_\mu=0$.

Finally, we may study the limit of further matter couplings according to the construction of section~\ref{subsec:matter}. From (\ref{LTMGlimit}) and (\ref{LTMSG}), we find that the relevant term in the matter Lagrangian (\ref{matter coupling}) is identified upon expansion
\begin{equation}
{\cal L}_{\rm matter} = \alpha\,{\cal L}^{\rm TMG}_{\rm matter} + {\cal O}(\alpha^2)
\quad
\Longrightarrow
\quad
T_{\mu\nu} = \alpha\,T_{\mu\nu}^{\rm TMG} + {\cal O}(\alpha^2)
\;.
\end{equation}
On the level of the field equations (\ref{eom_T}), and noting that
\be
\mathbb{S}_{\mu\nu}=\frac{3\lambda}{2}g_{\mu\nu}+S_{\mu\nu}
= -\frac{\mu^2}{\alpha^2}\,g_{\mu\nu} + {\cal O}\Big(\frac1\alpha\Big)
\;,
\ee
we find a contribution from the source tensor (\ref{defT}) 
\begin{equation} 
\tau\,\mathbb{T}_{\mu\nu} =
-\mu\,T_{\mu\nu}^{\rm TMG} 
+{\cal O}(\alpha)
\;.
\end{equation}

%%%%%%%%%%%%%%%%%%%%%%%%%%%%%%%%%%%%%%%%%%%%%%%%%%%%%%%
%%%%%%%%%%%%%%%%%%%%%%%%%%%%%%%%%%%%%%%%%%%%%%%%%%%%%%%

\section{(A)dS vacua}
\label{sec:AdSvacua}

%%%%%%%%%%%%%%%%%%%%%%%%%%%%%%%%%%%%%%%%%%%%%%%%%%%%%%%

We will now explore the landscape of (A)dS vacua of the MMSG model.  We will show that this model in general admits two maximally symmetric (A)dS/Minkowski vacua but only one of them preserves supersymmetry. We then compute the associated central charges in section 4.3. The linearization of the model around its AdS vacua is studied in section 4.4. Finally, in section 4.5 we determine the conditions on the parameters of the model such that there are no tachyons and ghosts and both central charges are positive for the AdS vacua. In particular, we localize the bulk/boundary unitary AdS vacua discovered in \cite{Bergshoeff:2014pca}. It turns out that the bulk/boundary clash can be avoided for both AdS vacua but not simultaneously.

Before analyzing the vacua of the model, let us recall that the fermionic couplings of the supersymmetric model are given in terms
of a parameter $\eta$ that is related to the parameters of the bosonic model by \eqref{lambda_eta0}, that is:
\begin{equation}
\lambda
=
\frac1{12}\left(\eta\,\tau+\tfrac1{\eta\,\kappa}\right)^2
-\frac{\tau}{3\,}\left(\eta \, \tau-\tfrac{1}{\eta\,\kappa}\right)
\;.
%\nonumber
\label{lambda_eta}
\end{equation}
Supersymmetrizability of the bosonic MMG model thus depends on the existence of real roots (for $\eta$) of this equation.
A necessary and sufficient set of conditions for the existence of such real roots is
\begin{align}
\bullet\;\;\;
&
{3\lambda}-\frac{\tau}{\kappa}+\tau^2 \ge 0 \;,
\nonumber\\
\bullet\;\;\;
&
\mbox{either}\;\;\,\frac{1}{\kappa\tau}\ge-1 \;\;\mbox{or}\;\;\,
\frac{3\lambda}{\tau^2} \ge -\frac{2}{\sqrt{-\kappa\tau}}
\;.
\label{cond_susy}
\end{align}
There are in general up to 4 real roots, pairwise related by
\begin{equation}
\eta \longrightarrow -\frac{1}{\eta\,\kappa\tau}
\;.
\label{flip_eta}
\end{equation}
Specifically, these are given by
\begin{equation}
{\eta_\pm} = 
(1\pm \Gamma) +  \sqrt{(1\pm \Gamma)^2+\frac1{\kappa\tau}}\;,
\qquad
\Gamma \coloneqq \sqrt{\frac{3\lambda}{\tau^2}-\frac{1}{\kappa\tau}+1}
\;,
\label{eta_Gamma}
\end{equation}
together with those obtained by the flip (\ref{flip_eta}),
which corresponds to flipping the sign in front of the square root in (\ref{eta_Gamma}).
Let us note that for the parameters $m$ and $\beta$ defined in (\ref{mbeta}), the flip (\ref{flip_eta})
amounts to
\begin{equation}
m\longrightarrow-m\;,\quad
\beta\longrightarrow\beta
\;,
\end{equation}
which leaves $\lambda$ invariant, while for the roots (\ref{eta_Gamma}) we find
\begin{equation}
\beta_\pm = \frac{1}2\,\Big(\eta_\pm\,\tau-\frac1{\eta_\pm\,\kappa}\Big) = (1\pm\Gamma)\,\tau
\;.
\label{betapm}
\end{equation}

In summary, a given bosonic MMG model thus can admit different supersymmetric extensions.
We will map out the landscape of these models in the following, together with their maximally symmetric vacua.

\subsection{Existence, location, and supersymmetry of vacua}

For maximally symmetric vacua where
$G_{\mu\nu} =  -\Lambda\,g_{\mu\nu}$,
the bosonic MMG field equations \eqref{MMGh} give a quadratic equation for the cosmological constant $\Lambda$
\be
\Lambda^2 + (4\tau^2 + 6\lambda) \Lambda \ + 9 \lambda^2 + \frac{4\tau^3}{\kappa} = (\Lambda +3\lambda)^2 + 4\tau^2 \left(\Lambda +\frac{\tau}{\kappa}\right)= 0 \, .
\label{cosmo}
\ee
The existence of real solutions for $\Lambda$ requires that
\bea
3\,\lambda-\frac{\tau}{\kappa}+\tau^2
&\ge& 0
\;.
\label{AdS_exist}
\eea
In terms of the original parameters of the MMG equation (\ref{MMG0}), this translates into the condition 
$\mu^2\bar{\sigma}^2 \geq \gamma \bar{\Lambda}_0$. 
Specifically, the values of the cosmological constant determined from (\ref{cosmo}) are given by
\begin{equation}
\Lambda=\Lambda_\pm = -\tau^2 \Big(
 (1\pm \Gamma)^2 + \frac{1}{\kappa\tau} \Big)
 \,,
 \label{Lpm}
\end{equation}
with $\Gamma\ge0$ defined in (\ref{eta_Gamma}), such that $\Lambda_+ \le\Lambda_-$. 
If the MMG model admits a supersymmetric extension, equation (\ref{AdS_exist}) is contained in (\ref{cond_susy}).
Every supersymmetric model thus possesses maximally symmetric vacua. Specifically, using the value of $\lambda$
given in \eqref{lambda_eta} the condition (\ref{AdS_exist}) is equivalent to $[\eta(\eta-2)\kappa\tau-1]^2 \geq 0$ and hence identically satisfied.
In this case we find
$\Gamma_{\textrm{MMSG}}=|\frac{\beta}{\tau}-1|$ where $\beta$ is defined in \eqref{mbeta} and we 
obtain the values of cosmological constants (\ref{Lpm}) as
\begin{equation}
\begin{array}{l}
\displaystyle{\Lambda_{\rm susy} =
 -\frac{(1+\eta^2\,\kappa\tau)^2}{4\,\eta^2\,\kappa^2}}
  \eqqcolon -\frac1{\ell_{\rm susy}^2}
\;,\\[2ex]
\displaystyle{\Lambda_{\rm ns} =
-\frac{1+\eta\,\kappa\tau[8+2\,\eta+\eta\,(\eta-4)^2\,\kappa\tau]}{4\,\eta^2\,\kappa^2}} 
\;,
\end{array}
\label{sns}
\end{equation}
with
\begin{equation}
\Lambda_{\rm susy} = \Lambda_\pm\;,\;\;
\Lambda_{\rm ns} = \Lambda_\mp
\qquad
\mbox{for}\quad
\frac{\beta}{\tau} \gtrless 1 
\;.
\end{equation}
Using the parameters introduced in \eqref{mbeta}, the cosmological constants (\ref{sns})  can be expressed as
\be 
%\Lambda_{\pm}= \frac{-(8\tau^2 + 12\lambda) \pm (8\tau^2 + 12\lambda -x^2)}{4} \Rightarrow
\Lambda_{\rm susy}= -m^2= -\beta^2- \frac{\tau}{\kappa} \,\,\, , \,\,\, \Lambda_{\rm ns}= -m^2 + 4\tau (\beta-\tau)=-\left(\beta -2\tau\right)^2- \frac{\tau}{\kappa}\, .
\label{cc}
\ee
Note that both, $\Lambda_{\rm susy}$ and $\Lambda_{\rm ns}$, are invariant under the flip (\ref{flip_eta}).
We also have the following useful identities
\begin{align}
\label{lambdalambda}
\Lambda_{\rm susy} + 3\lambda = -2\tau \beta \, \, , \, \,
\Lambda_{\mathrm{ns}} + 3 \lambda =2 \tau (\beta - 2 \tau)
\, \, , \, \,  3\lambda= \beta^2 -2\tau \beta + \frac{\tau}{\kappa}
\,.
\end{align}
The first vacuum in (\ref{sns}) is AdS (or Minkowski) and preserves {part of the} supersymmetry with the 
Killing spinor defined by
\begin{equation}
D[\omega]_\mu \epsilon 
-\frac1{2\ell_{\rm susy}}\,\gamma_\mu \epsilon =0 \, .
\end{equation}
{}From (\ref{susy_full}), it then follows that $\delta\psi_\mu=0$ is satisfied in the standard way for AdS (or Minkowski),
whereas $\delta \chi_\mu=0$ holds identically, as a consequence of \eqref{KS}. On the other hand, for 
the non-supersymmetric vacuum $\Lambda_{\rm ns}$  in (\ref{sns}), the Killing spinor equations for 
$\psi_\mu$ and $\chi_\mu$  \eqref{susy_full} cannot simultaneously be solved. The transformation 
\begin{equation}
    \beta \longrightarrow 2\tau-\beta \;,
    \label{betatr}
\end{equation}
leaves $\lambda$ the same while
interchanging supersymmetric and non-supersymmetric maximally symmetric vacua $\Lambda_{\rm susy} \longleftrightarrow \Lambda_{\rm ns}$ as can be seen from \eqref{cc} and \eqref{lambdalambda}. This corresponds to
switching $\eta_+\longleftrightarrow \eta_-$ from \eqref{betapm}.

\subsection{Parameter space of bosonic and supersymmetric models}
\label{subsec:parameterspace}

\begin{figure}[h]
   \centering
   \includegraphics[width=11cm]{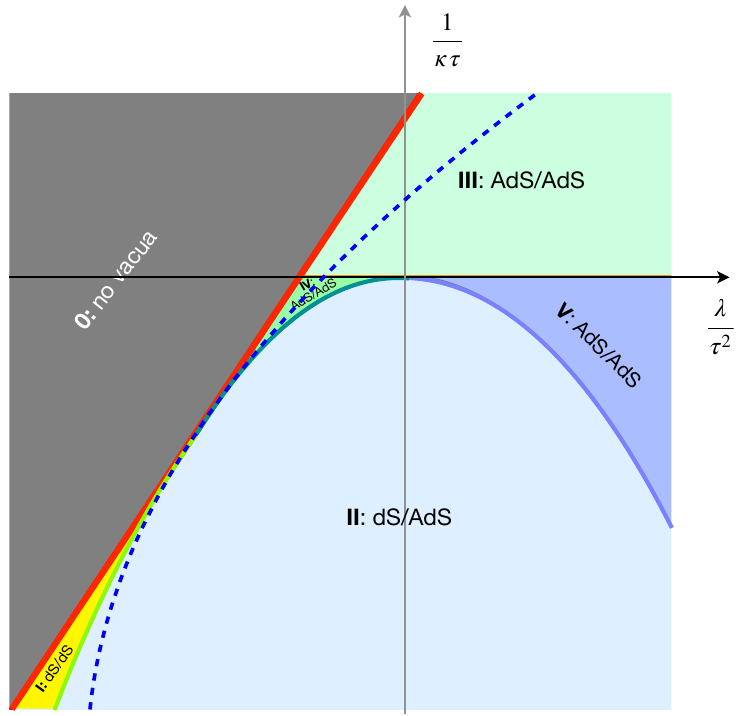}
   \caption{Different regions in the parameter space of bosonic models. 
   As long as the MMG model possesses an AdS vacuum,
   i.e.\ with the exception of the gray and yellow areas, it admits at least one supersymmetric extension.
   Along the parabola (\ref{parab}), there is always one Minkowski vacuum together with an (A)dS vacuum.
   The red line is the so-called merger line~\cite{Arvanitakis:2014yja}, $\Gamma=0$,  in which both (A)dS vacua of the model coincide.
   The dashed blue line is the chiral line (\ref{lchiral}) on which one of the central charges of the dual theory vanishes.
   The AdS vacua avoiding the bulk/boundary unitarity clash are all situated in region V, c.f.\ section~\ref{subsec:unitarity}.
}
   \label{fig:dSAdS}
\end{figure}

We can now map out the full landscape of bosonic and supersymmetric models and their vacuum structure. 
In Figure~\ref{fig:dSAdS}, we depict the parameter space of the bosonic models (\ref{LMMG}), parametrized
by the combinations $\frac{\lambda}{\tau^2}$ and $\frac1{\kappa\tau}$, both invariant under the scaling symmetry (\ref{scaling}).
Let us discuss the different regions separately.

\begin{enumerate}

\item[{\bf 0}:] 
This is the region in which
\begin{equation}
\frac{1}{\kappa\tau}
> 1+\frac{3\lambda}{\tau^2}
\;,
\end{equation}
i.e.\ condition (\ref{AdS_exist}) is not satisfied. Accordingly, the models in this region do not possess maximally symmetric vacua. 
Moreover, it follows from (\ref{cond_susy}) that these models do not admit supersymmetric extensions.

\item[{\bf I}:] 
This region is defined by
\begin{equation}
-\left(\frac{3\lambda}{2\tau^2}\right)^2 < \frac{1}{\kappa\tau} < {\rm min}\Big( 1+\frac{3\lambda}{\tau^2},-1\Big)
\,,\;\;\mbox{and}\;\;
\lambda<0
\;.
\end{equation}
According to (\ref{AdS_exist}), these models do admit maximally symmetric vacua, however it follows from (\ref{Lpm})
that both of these vacua are of dS type, i.e.\ $\Lambda_\pm>0$. Since the second condition of (\ref{cond_susy}) 
is violated, these models do not admit supersymmetric extensions.

\item[{\bf II}:] 
This region is defined by
\begin{equation}
\frac{1}{\kappa\tau} < -\left(\frac{3\lambda}{2\tau^2}\right)^2 
\;.
\end{equation}
In this region
\begin{equation}
(1- \Gamma)^2+\frac1{\kappa\tau}
< 0 < (1+ \Gamma)^2+\frac1{\kappa\tau}
\;,
\end{equation}
thus $\eta_-$ in (\ref{eta_Gamma}) is imaginary and only $\eta_+>0$ together with (\ref{flip_eta}) define
two supersymmetric extensions of the bosonic model. 
The cosmological constants (\ref{Lpm}) satisfy
\begin{equation}
\Lambda_+<0<\Lambda_-
\;,
\end{equation}
i.e.\ $\Lambda_+$ describes an AdS vacuum, and $\Lambda_-$ is dS. 
Comparing to (\ref{sns}) shows that the AdS vacuum $\Lambda_+$ is supersymmetric.

\item[{\bf III}:] This region is defined by
\begin{equation}
0 < \frac{1}{\kappa\tau} <  1+\frac{3\lambda}{\tau^2}\;,
\end{equation}
and  not covered by the Lagrangian of~\cite{Bergshoeff:2014pca}. 
All four roots of (\ref{lambda_eta}) are real and define supersymmetric extensions.
It follows from (\ref{Lpm}) that both vacua $\Lambda_\pm$ are of AdS type. 
Regarding supersymmetry, equations (\ref{sns}) show that  
\begin{equation}
\Lambda_{\rm susy}(\eta_\pm) = \Lambda_\pm = \Lambda_{\rm ns}(\eta_\mp)
\;,
\label{Lsns}
\end{equation}
for the supersymmetric extensions (\ref{eta_Gamma}), and likewise for those related by the flip (\ref{flip_eta}).
For each of the vacua $\Lambda_\pm$ of the bosonic model, there is thus a supersymmetric extension in which it is supersymmetric
while the other vacuum is non-supersymmetric.

\item[{\bf IV}:] 
This region is defined by
\begin{equation}
- {\rm min}\Big( 1\,,\Big(\frac{3\lambda}{2\tau^2}\Big)^2\Big) < \frac{1}{\kappa\tau} <  {\rm min}\Big(0\,,1+\frac{3\lambda}{\tau^2}\Big)
  \,,\;\;\mbox{and}\;\;
\lambda<0
\;.
\end{equation}
All four roots of (\ref{lambda_eta}) are real and positive and define supersymmetric extensions
with both vacua $\Lambda_\pm$ of AdS type. Equation (\ref{Lsns}) still holds and shows that
for each of the vacua $\Lambda_\pm$ of the bosonic model, there is a supersymmetric extension in which it is supersymmetric
while the other vacuum is non-supersymmetric.

\item[{\bf V}:] 
This region is defined by
\begin{equation}
- \Big(\frac{3\lambda}{2\tau^2}\Big)^2 < \frac{1}{\kappa\tau} < 0
  \,,\;\;\mbox{and}\;\;
\lambda>0
\;.
\label{region5}
\end{equation}
Again, all four roots of (\ref{lambda_eta}) are real, with $\eta_-<0<\eta_+$, and define supersymmetric extensions
with both vacua $\Lambda_\pm$ of AdS type. Equation (\ref{Lsns}) still holds and shows that
for each of the vacua $\Lambda_\pm$ of the bosonic model, there is a supersymmetric extension in which it is supersymmetric
while the other vacuum is non-supersymmetric.

\end{enumerate}
In particular, the analysis of the parameter space shows that every bosonic MMG model that possesses 
an AdS vacuum (i.e.\ lives within the regions II--V\ of Figure~\ref{fig:dSAdS}) admits at least two supersymmetric extensions 
(with parameters $\eta$ related by (\ref{flip_eta})) in which this vacuum is supersymmetric. 
In regions III--V\ where the model possesses two AdS vacua, there are different 
supersymmetric extensions, such that a given AdS vacuum is supersymmetric in one extension
and non-supersymmetric in another.

The models along the parabola
\begin{equation}
\frac{1}{\kappa\tau}=-\Big(\frac{3\lambda}{2\tau^2}\Big)^2
\;,
\label{parab}
\end{equation}
separating the different regions in Figure~\ref{fig:dSAdS}, all possess one Minkowski vacuum together with another maximally symmetric vacuum.
When $\Lambda_{\rm ns}=0$ on the parabola, 
one finds that $\eta>0$ is needed and moreover either $\eta \kappa\tau(\sqrt{\eta}-2)^2=-1$ or $\eta \kappa\tau(\sqrt{\eta}+2)^2=-1$. 
When $\Lambda_{\rm susy}=0$  we have
$1 + \eta^2 \kappa\tau=0$ which gives $\lambda=-2/(3\kappa^2\eta^3)$.

The red line in the figure is the so-called {\em merger} line~\cite{Arvanitakis:2014yja}
\begin{equation}
\frac{1}{\kappa\tau}
= 1+\frac{3\lambda}{\tau^2}\quad\Longleftrightarrow\quad
\Gamma=0
\;,
\label{merger}
\end{equation}
 in which both (A)dS vacua (\ref{Lpm}) of the model coincide. For $\frac\lambda{\tau^2}>-\frac23$ the vacuum is supersymmetric with
 \begin{equation}
\Lambda_{\rm susy}=\Lambda_{\rm ns}=    \Lambda_{\rm merger}=-\frac{\tau}{\kappa}
    (1+\kappa\tau) \,. 
\end{equation}
In particular, the model at the point
\begin{equation}
\frac1{\kappa\tau}=-1=1+\frac{3\lambda}{\tau^2}
\;,
\end{equation}
adjacent to regions 0, I, II, IV, has a single supersymmetric extension (with $\eta=1$) and a single supersymmetric Minkowski vacuum.
In this point, the parabola (\ref{parab}) meets the merger line (\ref{merger}) as well as the dashed 
{\em chiral} line (\ref{lchiral}) on which one of the central charges of the dual theory vanishes.
At this point, equations \eqref{param} give $\bar{\sigma}=\bar{\Lambda}_0=0$ for the parameters used in \cite{Bergshoeff:2014pca}.

Let us also note that the two regions $\kappa\tau < 0$ and $\kappa\tau > 0$ of the parameter space are not connected, 
since the model (\ref{LMasterEW}) degenerates for $\kappa\tau=0$, and $\frac1{\kappa\tau}=0$.

\begin{figure}[tb]
   \centering
   \includegraphics[width=16.6cm]{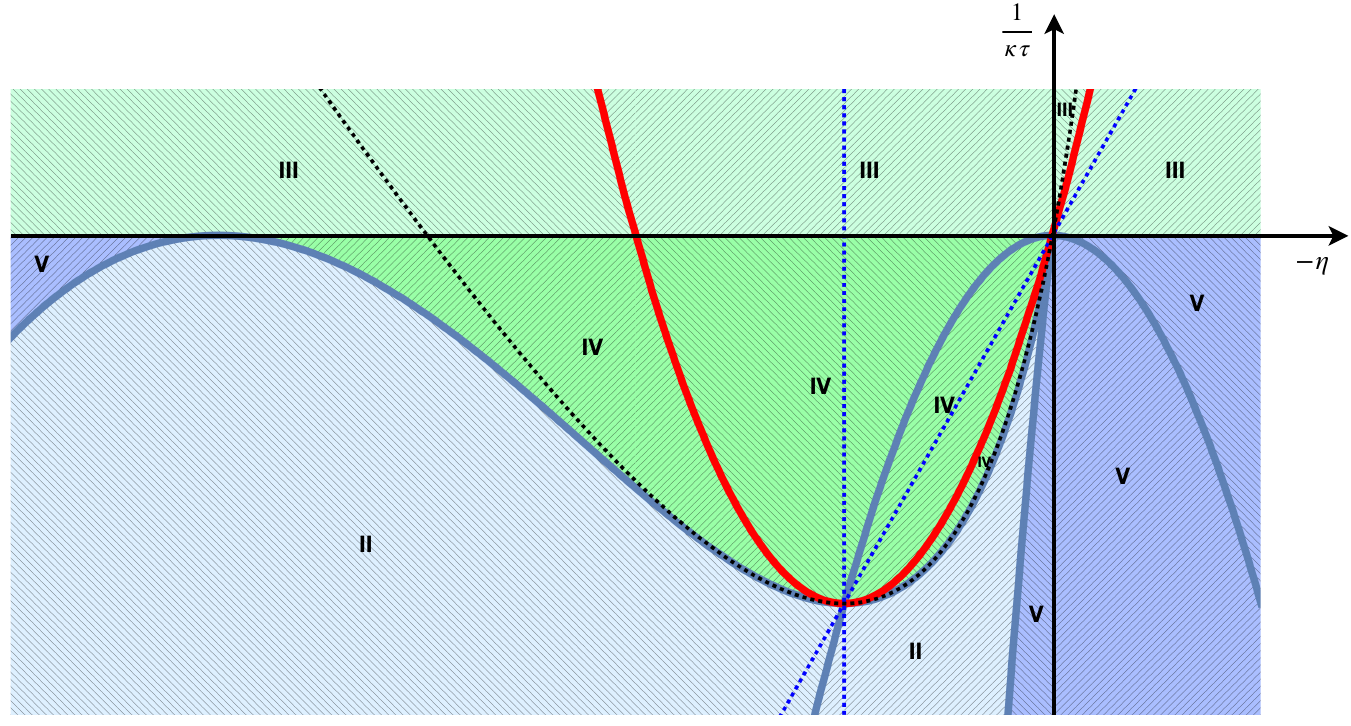}
   \caption{\footnotesize Different regions in the parameter space of supersymmetric models as a function of the parameter $\eta$. 
   The blue lines correspond to the parabola (\ref{parab})
   of Figure~\ref{fig:dSAdS}, separating the different regions.
   The red line is the merger line,  in which both (A)dS vacua of the model coincide.
   The dashed lines are the chiral lines (\ref{chirals1}), (\ref{chirals2}) on which one of the central charges of the dual theory vanishes.
   This map is a covering of part of Figure~\ref{fig:dSAdS} with different supersymmetric models mapping onto the same bosonic model
   as indicated for the different regions.
      The AdS vacua avoiding the bulk/boundary unitarity clash are all situated in region V, they are supersymmetric for $\eta<0$ and non-supersymmetric for $\eta>0$, c.f.\ section~\ref{subsec:unitarity}.
}
   \label{fig:landscape}
\end{figure}

In order to visualize the different supersymmetric extensions of a given bosonic model, it is instructive to plot the full landscape of 
supersymmetric models underlying the bosonic models of Figure~\ref{fig:dSAdS}. In Figure~\ref{fig:landscape} we depict the parameter space
of the supersymmetric MMSG Lagrangian (\ref{LMMSG}) parametrized by $\eta$ and $\frac1{\kappa\tau}$ both of which are invariant under the scaling symmetry \eqref{scaling}.
The different regions mapping onto the same bosonic model of Figure~\ref{fig:dSAdS}. are separated by the red merger line
(in which both AdS vacua of the model coincide) and the blue lines corresponding to the parabola (\ref{parab}).
Also, we note that the four quadrants of the parameter space in Figure~\ref{fig:landscape} are not connected 
since the model (\ref{LMMSG}) degenerates for $\frac1{\kappa\tau}=0$ and $\eta=0$.

\subsection{Central charges}

We now turn to the computation of the central charges for which we follow the method and formulas of \cite{Bergshoeff:2019rdb}. This requires starting from a bosonic Lagrangian expressed in terms of a spin connection $\Omega_\mu{}^a$, a vielbein $e_\mu{}^a$ and an auxiliary field $h_\mu{}^a$ such as (\ref{LMMGV3}) above.
Additionally, in \cite{Bergshoeff:2019rdb}, it is necessary to work with a torsionless connection, so we perform in (\ref{LMMGV3}) the shift $\Omega_\mu{}^a \rightarrow \Omega_\mu{}^a - \alpha h_\mu{}^a$, i.e.\ we go back to $\omega_\mu{}^a$ from (\ref{oOmega}). Examining the equations of motion of this new Lagrangian, for the AdS background with an Ansatz $h_\mu{}^a = c_h e_\mu{}^a$, one finds that the proportionality constant $c_h$ is determined by the equations of motion as
\begin{equation}
  c_h = \frac{1}{\alpha \tau} \left(\frac{\Lambda}{2} + \frac32 \lambda + \tau \theta \right).
\end{equation}
The remaining equations can be used to obtain the quadratic equation \eqref{cosmo} for $\Lambda$.
One can then apply the formula for the central charges $c_\pm$, given in \cite{Bergshoeff:2019rdb}, to find
\begin{equation}
  \label{eq:centralcharges}
  c_\pm = \left(-g_{e\Omega} - c_h g_{h\Omega} \pm \frac{1}{\ell} g_{\Omega\Omega} \right) \frac{3\ell}{2 G_3} \,,
\end{equation}
with $\ell=\frac{1}{\sqrt{-\Lambda}}>0$.
Here, $g_{e\Omega}$, $g_{h\Omega}$, $g_{\Omega\Omega}$ are the coefficients in front of the $2 \varepsilon^{\mu\nu\rho} e_{\mu}{}^a \partial_\nu \Omega_{\rho a}$, $2 \varepsilon^{\mu\nu\rho} h_{\mu}{}^a \partial_\nu \Omega_{\rho a}$ and $\varepsilon^{\mu\nu\rho} \Omega_{\mu}{}^a \partial_\nu \Omega_{\rho a}$ terms of the shifted Lagrangian. For completeness, we have also reintroduced the three-dimensional Newton constant $G_3$ which was set to 1 in the above. Putting everything together, one obtains\,\footnote{Taking the flat space $\ell \rightarrow \infty$ limit of $c_\pm$ is slightly subtle, as it naively diverges. In this regard, it is interesting to note that the limits $c_1 = \lim_{\ell \rightarrow \infty} (c_+ - c_-)$ and $c_2 = \lim_{\ell \rightarrow \infty} (c_+ + c_-)/\ell$ are well-defined. These limits reproduce the central charges of the asymptotic BMS-symmetry algebra of flat space TMG from those of the asymptotic Virasoro $\times$ Virasoro algebra of TMG around AdS$_3$ \cite{Bagchi:2012yk} and were also applied in studying the flat-space limit of MMG in \cite{Arvanitakis:2014xna}.}
\begin{equation}
  \label{eq:centralcharges2}
  \frac{2 G_3}{3 \ell} \,c_\pm =  \left(-1 + \frac{\kappa}{2\tau} \left(\Lambda + 3\lambda \right) \pm \frac{\kappa}{\ell} \right) \,. 
\end{equation}
This reproduces the central charges of \cite{Bergshoeff:2014pca}  (also given in eq.\ (5.21) of \cite{Bergshoeff:2019rdb}). 
For the supersymmetric AdS vacuum with the cosmological constant 
given by $\Lambda_{\rm susy}$ in \eqref{sns}, equation \eqref{eq:centralcharges2} simplifies to
\be\label{centralcharge}
\frac{2 G_3}{3 \ell}\,c_{\rm susy \,\pm} 
~=~ 
\left\{
\begin{array}{l}
\displaystyle{\frac{1}{\eta}-1}\\[1ex]
\displaystyle{-(1+\eta \kappa\tau)}
\end{array}
\right. \, .
\ee
The two central charges are exchanged under the flip (\ref{flip_eta}) of $\eta$. 
In the TMG limit \eqref{etatmg} the expressions (\ref{centralcharge}) 
approach the familiar expressions $\frac{2 G_3}{3 \ell}c_{\rm susy \,\pm} \rightarrow \alpha \left(\sigma \mp
\frac{m_0}{\mu}\right)$. The factor $\alpha$ is eliminated by multiplying the action with $\alpha^{-1}$ as 
seen in the limit \eqref{LTMGlimit}.

At the so-called {\it chiral points} one of the two central charges of the dual theory vanishes. Using 
\eqref{cosmo} and \eqref{eq:centralcharges2}, we see that this happens when
\begin{equation}
    \lambda_{\textrm{chiral}} = \frac{1}{12\kappa^2} + \frac{\tau}{2\,\kappa}-\frac{\tau^2}{4}
    \;,
    \label{lchiral}
\end{equation}
and the cosmological constant is given by
\begin{align} \label{eq:chiralpointcc}
  \Lambda_{\rm chiral}= -\frac{1}{4\kappa^2} \left(1+\kappa\tau\right)^2 = - (\beta-\tau)^2 \,.
\end{align}
These points lie on the dashed blue line in Figure~\ref{fig:dSAdS} which intersects with regions II--V.
For the supersymmetric AdS vacuum, using \eqref{cc}, \eqref{lambdalambda}, this condition implies that  
the parameter $\eta$ satisfies
\begin{equation}
(\eta-1)(\eta\,\kappa\,\tau+1)=0\;,
\label{chirals1}
\end{equation}
in accordance with (\ref{centralcharge}). On the other hand, for the non-supersymmetric AdS vacuum, one finds  
\begin{equation}
\eta(\eta-3)\,\kappa\,\tau=1 + \eta
\;.\label{chirals2}
\end{equation}
These chiral points lie on the curves depicted as dashed blue (\ref{chirals1}) and dashed black (\ref{chirals2}) lines in
Figure~\ref{fig:landscape}.

\subsection{The linearized spectrum}

In this section, we perform a linearized analysis of the MMSG model around an arbitrary AdS vacuum. This allows us to determine the spectrum of modes propagated by MMSG, as well as critical points in the parameter space, at which degeneracies in the mode spectrum occur. The linearization moreover  serves as the starting point for the next section, where we will identify the unitarity regions in parameter space, in which MMSG around AdS is both perturbatively stable and has an asymptotic symmetry algebra with positive central charges.

To linearize the MMSG Lagrangian \eqref{LMMSG} around an AdS vacuum with cosmological constant $\Lambda = -1/\ell^2$, we split the bosonic fields $e_\mu{}^{a}$, $\omega_\mu{}^a$ and $\varpi_\mu{}^a$ in background values $\tilde{e}_\mu{}^{a}$, $\tilde{\omega}_\mu{}^a$ and $\tilde{\varpi}_\mu{}^a$ and fluctuation fields $\mathrm{e}_\mu{}^{a}$, $\mathrm{w}_\mu{}^a$ and $\mathrm{v}_\mu{}^a$ in the following manner
\begin{align} \label{linAnsatz}
  e_{\mu}{}^{a} &= \tilde{e}_{\mu}{}^{a} + \varkappa \, \mathrm{e}_\mu{}^{a} \,, \qquad \omega_\mu{}^{a} = \tilde{\omega}_\mu{}^{a} + \varkappa \, \mathrm{w}_\mu{}^{a} \,, \qquad \varpi_\mu{}^{a} = \tilde{\varpi}_\mu{}^{a} + \varkappa\, \mathrm{v}_\mu{}^{a} \,.
\end{align}
Here, we have introduced a parameter $\varkappa$ to keep track of orders in fluctuation fields. We assume the background values of the fermionic fields to be zero, so that $\psi_\mu$ and $\chi_\mu$ are fields of order $\mathcal{O}(\varkappa)$ in what follows. In order not to overload the notation, we will denote the fluctuations of $\psi_\mu$ and $\chi_\mu$ with the same symbols, i.e., linearization replaces $\psi_\mu$ and $\chi_\mu$ by $\varkappa\, \psi_\mu$ and $\varkappa\, \chi_\mu$. Note that evaluating \eqref{KS} on the AdS background implies that
\begin{align}
  \tilde{\varpi}_\mu{}^{a} = \tilde{\omega}_\mu{}^{a} - \frac{1}{2 \tau} (\Lambda + 3 \lambda)\, \tilde{e}_\mu{}^{a} \,.
\end{align}
Expanding the Lagrangian \eqref{LMMSG} to $\mathcal{O}(\varkappa^2)$, one finds the following linearized Lagrangian
\begin{align}
  \mathcal{L}_{\rm lin} = \varkappa^2\,\mathcal{L}_{\rm lin,bos} + \varkappa^2\,\mathcal{L}_{\rm lin,ferm} \,,
\end{align}
with
\begin{align} \label{linbosLagr}
  \mathcal{L}_{\rm lin,bos} &= 2 \varepsilon^{\mu \nu \rho} \mathrm{e}_\mu{}^{a} D[\tilde{\omega}]_\nu \mathrm{w}_{\rho a} + \tau \varepsilon^{\mu \nu \rho} \mathrm{e}_\mu{}^{a} D[\tilde{\omega}]_\nu \mathrm{e}_{\rho a} + \kappa \varepsilon^{\mu\nu\rho} \mathrm{v}_\mu{}^{a} D[\tilde{\omega}]_\nu \mathrm{v}_{\rho a} \nonumber \\
                            & \quad + \varepsilon^{\mu\nu\rho} \varepsilon_{abc} \tilde{e}_{\mu}{}^{a} \mathrm{w}_\nu{}^{b} \mathrm{w}_\rho{}^{c} + \frac12 (3 \lambda - \Lambda) \varepsilon^{\mu\nu\rho} \varepsilon_{abc} \tilde{e}_\mu{}^{a} \mathrm{e}_\nu{}^{b} \mathrm{e}_\rho{}^{c} \nonumber \\
  & \quad + 2 \tau \varepsilon^{\mu\nu\rho} \varepsilon_{abc} \tilde{e}_\mu{}^{a} \mathrm{v}_\nu{}^{b} \mathrm{e}_\rho{}^{c} - \frac{\kappa}{2 \tau} (\Lambda + 3 \lambda) \varepsilon^{\mu\nu\rho} \varepsilon_{abc} \tilde{e}_\mu{}^{a} \mathrm{v}_\nu{}^{b} \mathrm{v}_\rho{}^{c} \,,
\end{align}
and 
\begin{align} \label{eq:linfermLagr}
  \mathcal{L}_{\rm lin,ferm} &= \frac1{\eta}\,\varepsilon^{\mu\nu\rho}
 \big[
( \bar\psi{}_\mu+\bar\chi_\mu) D[\tilde{\omega}]_\nu (\psi{}_\rho+\chi_\rho)
 -\eta\,\bar\psi{}_\mu D[\tilde{\omega}]_\nu \psi{}_\rho
 \big] + \frac12 \tau \varepsilon^{\mu\nu\rho} \bar{\chi}_\mu \tilde{\gamma}_\nu \chi_\rho \nonumber \\
                             & \quad - \frac{1}{4 \eta \tau} (\Lambda + 3 \lambda) \varepsilon^{\mu\nu\rho} \left(\bar{\psi}_\mu + \bar{\chi}_\mu\right) \tilde{\gamma}_\nu \left(\psi_\rho + \chi_\rho\right) - \tau \varepsilon^{\mu\nu\rho} \bar{\psi}_\mu \tilde{\gamma}_\nu \chi_\rho \nonumber \\
& \quad  + \frac{1}{4} \left( (\eta - 2) \tau + \frac{1}{\eta \kappa} \right) \varepsilon^{\mu\nu\rho} \bar{\psi}_\mu \tilde{\gamma}_\nu \psi_\rho \,,
\end{align}
where $\tilde{\gamma}_\mu \coloneqq \tilde{e}_\mu{}^a \gamma_a$.

\subsubsection{Bosonic spectrum}

At generic points in the parameter space of MMSG, one can redefine the bosonic fluctuation fields as
\begin{align}
  \label{eq:diagbosfluct}
  \mathrm{e}_{\mu a} &= f^{(-)}_{\mu a} + f^{(+)}_{\mu a} + p_{\mu a} \,, \nonumber \\
  \mathrm{w}_{\mu a} &= \frac{1}{\ell}\, f^{(-)}_{\mu a} -\frac{1}{\ell} \,f^{(+)}_{\mu a} 
  -\left(\tau + \frac{1}{2\tau}(\Lambda + 3 \lambda)\right) p_{\mu a} \,, \nonumber \\
  \mathrm{v}_{\mu a} &= -\frac{1}{\tau} \left( \frac12 \left(3 \lambda + \Lambda\right) - \frac{\tau}{\ell}\right) f^{(-)}_{\mu a} -\frac{1}{\tau} \left( \frac12 \left(3 \lambda + \Lambda\right) + \frac{\tau}{\ell}\right) f^{(+)}_{\mu a} - \frac{1}{\kappa} \,p_{\mu a} \,.
\end{align}
In terms of $f^{(\pm)}_{\mu a}$, $p_{\mu a}$, the bosonic part \eqref{linbosLagr} of the linearized Lagrangian then takes on the following diagonalized form
\begin{align} \label{linbosLagrgen}
  \mathcal{L}_{\mathrm{bos, lin}} &= \alpha_+\, \varepsilon^{\mu\nu\rho} f^{(+)}_{\mu}{}^a \left(D[\tilde{\omega}]_{\nu} f^{(+)}_{\rho a} -\frac{1}{\ell} \, \varepsilon_{abc}  f^{(+)}_{\nu}{}^b \tilde{e}_\rho{}^c  \right) \nonumber \\ & \ \ \ + \alpha_- \, \varepsilon^{\mu\nu\rho} f^{(-)}_{\mu}{}^a \left(D[\tilde{\omega}]_{\nu} f^{(-)}_{\rho a} + \frac{1}{\ell} \, \varepsilon_{abc} f^{(-)}_{\nu}{}^b \tilde{e}_{\rho}{}^c \right) \nonumber \\ & \ \ \ + \alpha_0 \, \varepsilon^{\mu\nu\rho} p_{\mu}{}^a \left( D[\tilde{\omega}]_{\nu} p_{\rho a} + M_p \, \varepsilon_{abc} p_{\nu}{}^b \tilde{e}_\rho{}^c \right) \,,
 \end{align}
where
\begin{align} \label{ccoeffs}
  \alpha_\pm \coloneqq& \left(\mp \frac{2}{\ell} + \tau + \frac{\kappa}{\tau^2} \left(\frac12 (\Lambda + 3 \lambda) \pm \frac{\tau}{\ell}\right)^2\right) \,, \qquad
%  \alpha_+ \coloneqq \left(-\frac{2}{\ell} + \tau + \frac{g}{\tau^2} \left(\frac12 (\Lambda + 3 \lambda) + \frac{\tau}{\ell}\right)^2\right) \,, \nonumber \\
  \alpha_0 \coloneqq \left(\frac{1}{\kappa} - \tau - \frac{1}{\tau} (\Lambda + 3 \lambda) \right) \,,
\end{align}
and
\begin{align} \label{eq:Mp}
  M_p \coloneqq  -\left(\tau + \frac{1}{2\tau}(\Lambda + 3 \lambda) \right) \,.
\end{align}
This shows that generically, the spectrum of bosonic modes of MMSG around an AdS vacuum consists of two massles modes $f^{(\pm)}_{\mu a}$ (with
$m\ell=\pm1$) and one massive mode $p_{\mu a}$ with mass $M_p$.
The coefficients $\alpha_-$ and $\alpha_+$ can be slightly simplified by using \eqref{cosmo} as
\be
\alpha_-= \frac{2\kappa}{\ell^2} + \frac{2}{\ell} - \frac{\kappa}{\tau \ell}\left(\Lambda + 3\lambda\right) \, , \, \qquad
\alpha_+= \frac{2\kappa}{\ell^2} - \frac{2}{\ell} + \frac{\kappa}{\tau \ell}\left(\Lambda + 3\lambda\right) 
 \, , %\, c_3 =   -\tau - \frac{1}{\tau} 
%(\Lambda + 3\lambda) + \frac{1}{g} \, ,
\label{generalcoefficient}
\ee
% with $\Lambda =-1/\ell^2$ and
% \be
% M= -\left(\tau + \frac{1}{2\tau}(\Lambda + 3\lambda) \right) \, ,
% \label{generalM}
% \ee
Note that from \eqref{generalcoefficient} using \eqref{cosmo} we then have
\begin{equation}
\alpha_+\alpha_-=
\frac{4\kappa}{\ell^2 \tau}\left(\tau^2 + (\Lambda + 3\lambda) - \frac{\tau}{\kappa}\right)= -\frac{4\kappa}{\ell^2}\alpha_0 \, .
\label{alphaalpha}
\end{equation}
At the chiral point \eqref{lchiral}, \eqref{eq:chiralpointcc}, the bosonic mass $M_p$, given in \eqref{eq:Mp}, is equal to $\sqrt{-\Lambda}$ or $-\sqrt{-\Lambda}$, such that the massive mode disappears from the spectrum. Moreover, the field redefinition \eqref{eq:diagbosfluct} is no longer invertible and with \eqref{alphaalpha} it follows that the coefficients in \eqref{ccoeffs} satisfy 
\begin{align} \label{coeffschiralpoint}
  \alpha_0 = 0 \qquad  \text{and} \qquad  \text{either } \ \alpha_+=0\ \text{ or }\  \alpha_- = 0 \,,
\end{align}
indicating that the linearized Lagrangian \eqref{linbosLagr} is no longer diagonalizable. In order to clarify the structure of the spectrum at the chiral point, we perform the following (invertible) field redefinition
\begin{align}
  \label{eq:bosfluctredefchiral}
  \mathrm{e}_{\mu a} &= 2 \kappa f_{\mu a} - 2 \kappa g^{(1)}_{\mu a} - 2 \kappa g^{(2)}_{\mu a} \,, \nonumber \\
  \mathrm{w}_{\mu a} &= (\kappa \tau + 1) f_{\mu a} + (\kappa \tau + 1) g^{(1)}_{\mu a} + g^{(2)}_{\mu a} \,, \nonumber \\
  \mathrm{v}_{\mu a} &= 2 \kappa \tau f_{\mu a} + 2 g^{(1)}_{\mu a} + 3 g^{(2)}_{\mu a} \,,
\end{align}
that brings the linearized Lagrangian \eqref{linbosLagr} into the form
\begin{align} \label{eq:linboslagrredefchiral}
\mathcal{L}_{\mathrm{bos, lin}} &=  4 \kappa (1 + \kappa \tau) \varepsilon^{\mu \nu \rho} g^{(1)}_\mu{}^a D[\tilde{\omega}]_{\nu} g^{(2)}_{\rho a} + (5 \kappa + 4 \kappa^2 \tau) \varepsilon^{\mu \nu \rho} g^{(2)}_\mu{}^a D[\tilde{\omega}]_{\nu} g^{(2)}_{\rho a} \nonumber \\ & \ \ \ -2 (1 + \kappa \tau)^2 \varepsilon^{\mu \nu \rho} \varepsilon_{abc} \tilde{e}_{\mu}{}^{a} g^{(1)}_{\nu}{}^b g^{(2)}_{\rho}{}^{c} - \frac12 \left(5 + 7 \kappa \tau + 2 \kappa^2 \tau^2 \right) \varepsilon^{\mu \nu \rho} \varepsilon_{abc} \tilde{e}_{\mu}{}^{a} g^{(2)}_{\nu}{}^b g^{(2)}_{\rho}{}^c \nonumber \\ & \ \ \ + 2 \kappa (1 + \kappa \tau)^2 \varepsilon^{\mu\nu\rho} f_{\mu}{}^a \left( 2 D[\tilde{\omega}]_{\nu} f_{\rho a} + \left(\tau + \frac{1}{\kappa}\right) \, \varepsilon_{abc}  f_{\nu}{}^b \tilde{e}_\rho{}^c  \right) .
\end{align}  
The linearized equations of motion that follow from this Lagrangian are then given by
\begin{align} \label{eq:chiralpointeoms}
  & \mathcal{D}_{[\mu} f_{\nu] a}  = 0 \,, \qquad  \bar{\mathcal{D}}_{[\mu} g^{(2)}_{\nu] a}  = 0 \,, \qquad  \bar{\mathcal{D}}_{[\mu} g^{(1)}_{\nu]a}  + \frac12 \tau \varepsilon_{abc} \tilde{e}_{[\mu}{}^b g^{(2)}_{\nu]}{}^c = 0 \,,
\end{align}
where we have introduced differential operators $\mathcal{D}$ and $\bar{\mathcal{D}}$ that act on fluctuations $f_{\mu a}$ and $g_{\mu a}$ as follows
\begin{align}
  & \mathcal{D}_{[\mu} f_{\nu] a} \coloneqq D[\tilde{\omega}]_{[\mu} f_{\nu] a} + \frac12 \left(\tau + \frac{1}{\kappa}\right) \varepsilon_{abc} \tilde{e}_{[\mu}{}^b f_{\nu]}{}^c \,, \nonumber \\ & \bar{\mathcal{D}}_{[\mu} g_{\nu] a} \coloneqq D[\tilde{\omega}]_{[\mu} g_{\nu] a} - \frac12 \left(\tau + \frac{1}{\kappa}\right) \varepsilon_{abc} \tilde{e}_{[\mu}{}^b g_{\nu]}{}^c \,.
\end{align}
Since $\left(\tau + 1/\kappa\right) = \pm 2\sqrt{-\Lambda}$, see \eqref{eq:chiralpointcc}, the first two equations of \eqref{eq:chiralpointeoms} imply that $f_{\mu a}$ and $g^{(2)}_{\mu a}$ correspond to massless modes. Acting on the fluctuation $g^{(1)}_{\mu a}$ with the operator $\bar{\mathcal{D}}$ gives the mode $g^{(2)}_{\mu a}$ that is itself annihilated by $\bar{\mathcal{D}}$. By definition, $g^{(1)}_{\mu a}$ is then a so-called logarithmic mode associated to the massless mode $g^{(2)}_{\mu a}$. Its presence indicates that the dual CFT of MMSG at the chiral point \eqref{lchiral} is a non-unitary logarithmic CFT \cite{Grumiller:2008qz,Grumiller:2008pr,Grumiller:2008es,Maloney:2009ck,Skenderis:2009nt,Grumiller:2013at, Deger:2018pzj}.

In the following, we will turn our attention to the spectrum of fermionic modes. We will distinguish between two cases, according to whether the AdS vacuum under consideration is supersymmetric or not. We will be interested in investigating at which points in parameter space MMSG can be unitary. Since we just saw that at the chiral point \eqref{lchiral}, MMSG can not be described by a unitary CFT, we will not consider what happens at the special points of section \ref{subsec:parameterspace}.

\subsubsection{Supersymmetric vacua}

The linearized fermionic Lagrangian \eqref{eq:linfermLagr} around a supersymmetric AdS vacuum can be written as
\begin{align}
  \mathcal{L}_{\rm lin,ferm, susy} &= \frac1{\eta}\,\varepsilon^{\mu\nu\rho}
 \big[
( \bar\psi{}_\mu+\bar\chi_\mu) D[\tilde{\omega}]_\nu (\psi{}_\rho+\chi_\rho)
 -\eta\,\bar\psi{}_\mu D[\tilde{\omega}]_\nu \psi{}_\rho
 \big] + \frac{m}{\eta} \varepsilon^{\mu\nu\rho} \bar{\psi}_\mu \tilde{\gamma}_\nu \chi_\rho \nonumber \\ & \quad - 
                           \frac{m}{2 \eta} (\eta - 1) \varepsilon^{\mu\nu\rho} \bar{\psi}_\mu \tilde{\gamma}_\nu \psi_\rho  + \left(\frac{\beta}{ \eta}-\frac{m}{2 \eta}  \right) \varepsilon^{\mu\nu\rho} \bar{\chi}_\mu \tilde{\gamma}_\nu \chi_\rho \,.
\end{align}
Assuming that $\eta \neq 1$ (which is a chiral point which we discussed above), this Lagrangian can be brought into the following diagonalized form
  \begin{align} \label{linfermsusy}
  \mathcal{L}_{\mathrm{lin,ferm,susy}} &= \left(\eta^{-1} - 1\right) \varepsilon^{\mu\nu\rho} \bar{\rho}_{\mu\, 1} \left( D[\tilde{\omega}]_\nu \rho_{\rho\, 1} + \frac{m}{2} \,\tilde{\gamma}_\nu \rho_{\rho\, 1} \right) \nonumber \\ & \qquad + \left(\eta - 1\right) \varepsilon^{\mu\nu\rho} \bar{\rho}_{\mu\, 2} \left[ D[\tilde{\omega}]_\nu \rho_{\rho\, 2} - \left(\tau + \frac{m}{2} - \beta \right) \tilde{\gamma}_\nu \rho_{\rho\, 2} \right] ,
  \end{align}
  where $\rho_{\mu\, 1}$ and $\rho_{\mu\, 2}$ are defined via the following field redefinitions
 \begin{align}
  \psi_\mu = \rho_{\mu\, 1} + \rho_{\mu\, 2} \,, \qquad \qquad \qquad 
  \chi_\mu = \left(\eta - 1\right) \rho_{\mu\, 2} \,.
 \end{align}
 Since $1/|m|$ is equal to the AdS length $\ell_{\mathrm{susy}}$ of the supersymmetric vacuum, it follows that $\rho_{\mu\, 1}$ corresponds to a massless spin-$3/2$ mode. On the other hand, $\rho_{\mu\, 2}$ is a massive spin-$3/2$ mode, with mass parameter $M_{\rho_2}$ given by
 \begin{align}
   M_{\rho_2} \coloneqq \tau + \frac{m}{2} - \beta =  -\frac{m}{2 \eta} + \frac{(1-\eta)}{\eta} \left(\beta -\frac{m}{2}\right) .
 \end{align}
 For $\Lambda = \Lambda_{\mathrm{susy}}$, the bosonic mass $M_p$ \eqref{eq:Mp} can be expressed as
 \begin{align}
  M_p = 
-\frac{1-\eta\,(\eta-2)\,\kappa\,\tau}{2\,\eta\,\kappa} \,. 
\end{align}
Together, the system has one massive bosonic and one massive fermionic mode, with masses related by
 \begin{align}
M_{\rho_2}\,\ell_{\mathrm{susy}} &= M_p\,\ell_{\mathrm{susy}} -\frac12 \,,
\label{masses_susy}
 \end{align}
 in accordance with supersymmetry. For future reference, we also note that the coefficients \eqref{ccoeffs} for the supersymmetric vacuum take a nicely factorized form
\begin{align}
  \alpha_+ &= \eta^{-2}\,(1-\eta)\,\kappa^{-1}\,(1+\eta^2\,\kappa\,\tau) \,, \qquad \quad 
  \alpha_- = \eta^{-1} \,\kappa^{-1}\,(1+\eta\,\kappa\,\tau)(1+\eta^2\,\kappa\,\tau) \,, \nonumber \\
  \alpha_0 &= - \eta^{-1}\,(1-\eta)\,\kappa^{-1}\,(1+\eta\,\kappa\,\tau) \,.
\end{align}

\subsubsection{Non-supersymmetric vacua}

In the generic case, where $\Lambda_{\mathrm{ns}} \neq 0$, the field redefinitions
\begin{align}
  \psi_{\mu} &= \tau \left(\tilde{\rho}_{\mu\, 1} + \tilde{\rho}_{\mu\, 2}\right) \,, \nonumber \\
  \chi_{\mu} &= \left(\frac12 \sqrt{-\Lambda_{\mathrm{ns}}}-\frac{m}{2} \right) \tilde{\rho}_{\mu\, 1} - \left(\frac{m}{2} + \frac12 \sqrt{-\Lambda_{\mathrm{ns}}}\right) \tilde{\rho}_{\mu\, 2} \,,
\end{align}
bring the linearized fermionic Lagrangian \eqref{eq:linfermLagr} to the diagonalized form
 \begin{align} \label{linfermnonsusy}
  & \mathcal{L}_{\mathrm{lin,ferm,ns}} = \eta^{-1}\left(-\frac{\Lambda_{\mathrm{ns}}}{2} + \left(  \tau -\frac{m}{2} \right)  \sqrt{-\Lambda_{\mathrm{ns}}} \right) \varepsilon^{\mu\nu\rho} \bar{\tilde{\rho}}_{\mu\, 1} \left( D[\tilde{\omega}]_\nu \tilde{\rho}_{\rho\, 1} + \frac{1}{2} \sqrt{-\Lambda_{\mathrm{ns}}}\tilde{\gamma}_\nu \tilde{\rho}_{\rho\, 1} \right) \nonumber \\ & \qquad + \eta^{-1}\left(-\frac{\Lambda_{\mathrm{ns}}}{2} - \left(\tau -\frac{m}{2} \right)  \sqrt{-\Lambda_{\mathrm{ns}}}\right) \varepsilon^{\mu\nu\rho} \bar{\tilde{\rho}}_{\mu\, 2} \left[ D[\tilde{\omega}]_\nu \tilde{\rho}_{\rho\, 2} - \frac12 \sqrt{-\Lambda_{\mathrm{ns}}} \tilde{\gamma}_\nu \tilde{\rho}_{\rho\, 2} \right] .
  \end{align}
This shows that the fermionic spectrum of MMSG around a non-supersymmetric AdS vacuum consists of two massless spin-$3/2$ modes.
Unlike (\ref{masses_susy}) they no longer form a multiplet with the bosonic mode of mass \eqref{eq:Mp}.

\subsection{Unitarity analysis}
\label{subsec:unitarity}

We finally analyze the conditions for absence of tachyons and ghosts as well as for positivity of the central charges for the AdS vacua 
of our theory to identify the region in which the clash of bulk-boundary unitarity is avoided and the interplay with supersymmetry.
All these unitarity conditions are properties of the bosonic model, i.e.\ can be characterized by the parameters $\kappa$, $\tau$, and $\lambda$
of the Lagrangian (\ref{LMMG}).
The parameter $\eta$ selecting the supersymmetric extension on the other hand encodes the fact if the AdS vacuum is supersymmetric or not.

For the general AdS vacuum, using the value of $M_p$ given in \eqref{eq:Mp}, the no-tachyon condition $M_p^2\ell^2>1$ takes the form
\be
\tau^2 + (\Lambda + 3\lambda) - \frac{\tau}{\kappa} >0 \, .
\label{no-tachyon2}
\ee
According to the discussion of \cite{Bergshoeff:2014pca} the no-ghost condition is given as $\alpha_0 M_p<0$, with $\alpha_0$ from (\ref{ccoeffs}). 
With \eqref{eq:Mp}, this condition becomes
\be
- \left(\tau^2 + (\Lambda + 3\lambda) - \frac{\tau}{\kappa}\right)\left(1+ \frac{1}{2\tau^2}(\Lambda + 3\lambda)\right) >0 \,.
\label{noghostgeneral}
\ee
Combining this inequality with the general no-tachyon condition \eqref{no-tachyon2}
we find
\be
-\left(1+ \frac{1}{2\tau^2}(\Lambda + 3\lambda)\right) >0
\,.
\label{no-ghost2}
\ee
For the analysis of the positivity of the central charges, 
comparing \eqref{eq:centralcharges2} with \eqref{generalcoefficient} we see that
\be 
\frac{2 G_3}{3 \ell} c_\pm=\pm \frac{\ell}{2}\alpha_{\pm}\,.
\label{centralalpha}
\ee
Since requiring both central charges to be positive implies $-\alpha_+\alpha_->0$, we deduce from the no-tachyon condition \eqref{no-tachyon2} 
together with \eqref{alphaalpha} that unitarity always requires $\kappa\tau<0$.

Finally, the analysis of unitarity is sensitive to changing the action by an overall minus sign. Whereas 
the no-tachyon condition \eqref{no-tachyon2} remains invariant under this change, this will induce
a minus sign in the no-ghost condition \eqref{no-ghost2} and also in the central charges \eqref{eq:centralcharges2}. 
Note that, $\kappa\tau<0$ is still required since this follows from imposing the no-tachyon condition and the product of central charges to be positive, both of which are unaffected by the sign change of the action.

As for the special points, note that the no-tachyon condition \eqref{no-tachyon2} is violated at the chiral points \eqref{lchiral}. Meanwhile,
at the merger line we have $\beta = \tau$ which implies $\Lambda+ 3\lambda=-2\tau^2$ which violates the no-ghost condition  \eqref{noghostgeneral}. The situation for Minkowski vacua is more subtle and requires further investigation as discussed above.

After these general remarks, we now proceed to a detailed analysis of unitarity conditions and their interplay with supersymmetry. 
The analysis of section~\ref{subsec:parameterspace} has shown that every AdS vacuum is supersymmetric in some supersymmetric extension,
i.e.\ for some choice of $\eta$. Without loss of generality we may thus start with the supersymmetric  AdS vacuum (\ref{sns}). 
In this case, the no-tachyon condition \eqref{no-tachyon2} takes the factorized form
\bea
\eta\,(1-\eta)\,\kappa\,\tau\,(1+\eta\,\kappa\,\tau) &>& 0
\;.
\label{no-tachyon}
\eea
It turns out that the sign of $\eta$ is crucial for the rest of this analysis and only one choice works. 
Let us first assume $\eta<0$ or equivalently $\tau\beta<0$.\footnote{As mentioned above, this region of parameter space was not considered in 
the analysis of \cite{Deger:2022gim} where $\eta$ was represented as $\eta=\zeta^2$.}
In this case, the no-tachyon condition (\ref{no-tachyon}) is identically satisfied as a consequence of $\kappa\,\tau<0$.
Positivity of the central charges \eqref{centralcharge} requires an overall minus sign in front of the action 
and furthermore imposes the conditions
\begin{equation}
1-\eta >0 \;,\hspace{0.5cm} \textrm{and} \hspace{0.5cm}   1+\eta \kappa\tau>0 \, . 
\end{equation}
These conditions are also satisfied with our assumptions $\eta<0$ and $\kappa \tau<0$. 
Finally, with the overall sign in the action,
the condition for absence of ghosts \eqref{noghostgeneral} becomes
\be
(1+\eta \kappa\tau) - \eta (\eta-1)\kappa\tau >0
\;,
\label{no-ghost}
\ee
which also is identically satisfied, and hence does not further restrict the parameters.
It is easy to show that the other choice for $\eta$, namely $\eta>0$, cannot avoid the clash. 
To summarize, all supersymmetric AdS vacua in the region $\eta<0$ and $\kappa \tau<0$,
i.e.\ within the full lower-right quadrant of Figure~\ref{fig:landscape}, are bulk and boundary unitary.
In the bosonic model, these are the AdS vacua situated in region V of Figure~\ref{fig:dSAdS},
with cosmological constant $\Lambda_{\rm susy}(\eta_-)=\Lambda_-$, c.f.\ (\ref{Lsns}).
These vacua have first been found in \cite{Bergshoeff:2014pca}.
In the bosonic parameter space, the unitarity region is thus defined by (\ref{region5})
\begin{equation}
- \Big(\frac{3\lambda}{2\tau^2}\Big)^2 < \frac{1}{\kappa\tau} < 0
  \,,\;\;\mbox{and}\;\;
\lambda>0
\;.
\label{region5b}
\end{equation}
In terms of the TMG parameters \eqref{rule12} (and with the choice $1+\alpha\sigma>0$), 
these conditions translate into  
\begin{equation}
    \,\Lambda_0<\frac{4\mu^2\,(1+\alpha\sigma)^3}{\alpha^3} \, .
\end{equation}
Moreover, the condition of an overall minus sign in front of the action is equivalent to the condition $\alpha<0$ 
(or equivalently $\gamma>0$ from \eqref{param} and \eqref{rule12}) 
found by multiplying our action with $\alpha^{-1}$, which is necessary to get a nonzero action in the TMG limit \eqref{LTMGlimit}.
This is precisely consistent with the findings of \cite{Bergshoeff:2014pca, Arvanitakis:2014xna} (see, in particular equation (60) of \cite{Arvanitakis:2014xna}).

According to the analysis of section~\ref{subsec:parameterspace}, 
these unitary AdS vacua also arise as non-supersymmetric vacua in the supersymmetric models in 
region V with parameter $\eta_+>0$. 
These are the models in regions V in the lower-left quadrant of Figure~\ref{fig:landscape}.

To summarize, all AdS vacua in regions II--IV suffer from the bulk/boundary unitarity clash. In contrast, all models in region V  (\ref{region5b})
admit an AdS vacuum that evades the unitarity clash as found in \cite{Bergshoeff:2014pca}. 
This vacuum is supersymmetric in the extension with $\eta=\eta_-<0$ from \eqref{eta_Gamma},
and non-supersymmetric in the extension with $\eta=\eta_+>0$.

%%%%%%%%%%%%%%%%%%%%%%%%%%%%%%%%%%%%%%%%%%%%%%%%%%%%%%%
%%%%%%%%%%%%%%%%%%%%%%%%%%%%%%%%%%%%%%%%%%%%%%%%%%%%%%%

\section{Conclusions}

The starting point for this work, which details the findings of \cite{Deger:2022gim}, has been the new and universal action principle \eqref{LMasterEW} to build third-way consistent gravitational field equations presented in section \ref{sec:bosonic}. This variational principle allows to recover known models such as e.g. minimal massive gravity (MMG), exotic massive gravity, and exotic general massive gravity, but also new models which were missed in \cite{Ozkan:2018cxj} and which are of increasingly higher order. Apart from this possibility of constructing new bosonic models, the advantage of our action principle is to allow for a systematic description of the matter couplings. In particular, the coupling to fermionic matter has enabled us to obtain a supersymmetric extension (up to and including quartic order in the fermions) of minimal massive gravity (MMSG) in section \ref{sec:susy}. It is the first time that a third-way consistent gravity model is made supersymmetric, and as a byproduct of the construction topologically massive supergravity can also be recovered.

In section \ref{sec:AdSvacua} we have performed a detailed analysis of the AdS vacua of the new MMSG model. This establishes that whenever a bosonic MMG model has an AdS vacuum then it admits up to four supersymmetric extensions. For all the AdS vacua we have computed the (bosonic and fermionic) mass spectra and the central charges, key to the discussion of unitarity. We have summarized the regions of parameter space in which the unitary vacua are compatible or not with supersymmetry on figures \ref{fig:dSAdS} and \ref{fig:landscape}.

There are many interesting directions to explore, having to do on the one hand with the bosonic theories alone, and on the other hand with their supergravity features. A few interesting prospects are as follows:
   
\begin{itemize}

\item Having established the supersymmetric extension of MMG, it would be interesting to construct and classify its supersymmetric BPS solutions. This investigation should be facilitated by the observation that after eliminating the auxiliary fields, i.e.\ at the level of the metric field equations, the supersymmetry transformations of MMSG are the same as those of super TMG \eqref{susytmg} (with the constant $m_0$ replaced by $m$). With the Killing spinors of both theories of the same form,  it would be interesting to compare with the results of \cite{Gibbons:2008vi}.

\item
The construction should allow an extension to supersymmetric matter couplings, including ${\cal N}=1$ scalar and vector multiplets. It would be interesting to investigate how further couplings affect the unitarity analysis presented for the minimal model.

\item
A challenging task would be the supersymmetrization of the higher order extensions of the model \cite{Ozkan:2018cxj,Afshar:2019npk}.
The action (\ref{LMasterEW}) which naturally accommodates all such generalizations provides a natural starting point.
A universal construction of such third-way consistent supersymmetric models would presumably require a
superspace formulation.

\item We have identified (up to four) different minimal supersymmetric  extensions of the same bosonic MMG in which different vacua of the bosonic model appear supersymmetric. This points at an underlying structure of extended supersymmetry into which these models could be embedded and recovered by different truncations, as in the so-called twin supergravities \cite{Roest:2009sn}. As a technical challenge this would require to embed the single massive spin-2 degree of freedom into some extended multiplet. More generally, this raises the question of whether there exists a possible upper bound for the number of supercharges $\mathcal{N}$, like in the Yang-Mills case which does not allow ${\cal N}$ > 1 \cite{Deger:2022znj}.

\item Relatedly, one could try to understand the presence of the massive spin-$3/2$ mode in MMSG as an effect of
spontaneous breaking of a local symmetry of the Chern-Simons theory based on an
appropriate superalgebra in analogy to the MMG case \cite{Chernyavsky:2020fqs}.

\end{itemize}

We hope to come back to these points in the future.

%%%%%%%%%%%%%%%%%%%%%%%%%%%%%%%%%%%%%%%%%%%%%%%%%%%%%%%

\section*{Acknowledgements}

We would like to thank Olaf Hohm and Mehmet Ozkan for useful discussions and comments. 
JR and NSD are grateful to the Erwin Schr\"odinger Institute (ESI), Vienna where part of this work was done in the framework of the ``Research in Teams'' Programme. NSD would like to thank ICTP and its HECAP research section for hospitality and financial support during some part of this paper. In the last stages, this work was supported by the Croatian Science Foundation project IP-2022-10-5980 ``Non-relativistic supergravity and applications''.

% \bibliographystyle{utphys}
% \bibliography{refs}

\providecommand{\href}[2]{#2}\begingroup\raggedright\endgroup

\end{document}